\documentclass[times,sort&compress,3p]{elsarticle}
\usepackage[labelfont=bf]{caption}

\usepackage{amsmath,amsfonts,amssymb,amsbsy,amsthm,booktabs,color,epsfig,graphicx,hyperref,url}
\usepackage{bm}
\usepackage{hyperref}
\hypersetup{colorlinks,linkcolor={blue},citecolor={blue},urlcolor={blue}}  
\usepackage{float}
\usepackage[flushleft]{threeparttable}
\usepackage{enumerate}
\usepackage{natbib}
\usepackage[dvipsnames]{xcolor}
\usepackage{ulem}
\usepackage{subcaption}
\usepackage[ruled,linesnumbered]{algorithm2e}
\usepackage{array}

\usepackage{booktabs}
\usepackage{multirow}

\theoremstyle{plain}
\newtheorem{theorem}{Theorem}

\newtheorem{lemma}{Lemma}

\theoremstyle{definition}

\newcommand{\dif}{{\mbox{d}}}
\newcommand{\sgn}{{\mbox{sign}}}

\newcommand{\bmT}{{\bm T}}
\newcommand{\bmY}{{\bm Y}}

\newcommand{\bmX}{{\bm X}}

\def\bSig\mathbf{\Sigma}

\newcommand{\PE}{\mbox{PE}}
\newcommand{\lambdaCV}{\lambda_n^{ \mbox{CV}}}
\newcommand{\lambdaSE}{\lambda_n^{1 \mbox{SE}}}
\begin{document}

\begin{frontmatter}

\title{Variable Selection for Stratified Sampling Designs in Semiparametric Accelerated Failure Time Models with Clustered Failure Times}

  \author[1,2]{Ying Chen}
  \author[1]{Chuan-Fa Tang\corref{mycorrespondingauthor}}
  \author[3]{Sy Han Chiou}
  \author[1]{Min Chen}

  \address[1]{Department of Mathematical Sciences, University of Texas at Dallas, Richardson, TX 75080, USA}
  \address[2]{Department of Epidemiology, Harvard T.H. Chan School of Public Health, Boston, MA 02115, USA}
  \address[3]{Department of Statistics and Data Science, Southern Methodist University, TX 75275, USA}

  \cortext[mycorrespondingauthor]{Corresponding author. Email address: \url{chuan-fa.tang@utdallas.edu}}

\begin{abstract}
  In large-scale epidemiological studies, statistical inference is often complicated by high-dimensional covariates under stratified sampling designs for failure times.
    Variable selection methods  developed for full cohort data do not extend naturally to stratified sampling designs, and appropriate adjustments for the sampling scheme are necessary. Further challenges arise when the failure times are clustered and exhibit within-cluster dependence. As an alternative of Cox proportional hazards (PH) model when the PH assumption is not valid, the penalized Buckley-James (BJ) estimating method for accelerated failure time (AFT) models can potentially handle within-cluster correlation in such setting by incorporating generalized estimating equation (GEE) techniques, though its practical implementation remains hindered by computational instability. We propose a regularized estimating method within the GEE framework for stratified sampling designs, in the spirit of the penalized BJ method but with a  reliable inference procedure.  
    We establish the consistency and asymptotic normality of the proposed estimators and
    show that they achieve the oracle property.
    Extensive simulation studies demonstrate that our method outperforms existing methods
    that ignore sampling bias or within-cluster dependence.
    Moreover, the regularization scheme effectively selects relevant variables even with moderate sample sizes.
    The proposed methodology is illustrated through applications to a dental study. 
\end{abstract}


\end{frontmatter}

\section{Introduction\label{sec:1}}

Stratified sampling designs are widely used in large-scale epidemiological studies,
wherein the population is partitioned into mutually exclusive subgroups, or strata,
and subjects are randomly selected within each stratum.
The selection probabilities are typically designed to ensure adequate representation of key subpopulations,
thereby enhancing the efficiency and generalizability of the study.
As a special case of stratified sampling,
stratified case-cohort sampling is a common variation of the case-cohort design \citep{prentice1986case},
frequently used in failure time studies, particularly when disease incidence is low and covariate measurement is costly.
An example of this design appears in a retrospective dental study \citep{caplan2005root} analyzed in Section 6,
where the time to extraction of root canal (RC)-filled teeth is the event of interest, yet only a small fraction of patients underwent such extractions.
To construct the case-cohort sample used in the study, a random sample was drawn separately from two strata,
namely ``cases'' (patients who underwent extraction of RC-filled teeth) and ``controls'' (patients who did not).
Although case patients represented only a small fraction of the overall population, their inclusion was crucial for addressing the study objectives.
To ensure sufficient statistical power, unequal sampling probabilities were applied, with 85\% of case patients and 11\% of control patients selected.
This sampling scheme reduces costs and effort by collecting covariates only within the stratified subcohort,
but the use of unequal selection probabilities can introduce statistical bias.
Additionally, each sampled patient contributed both an RC-filled tooth and a contralateral non–RC-filled tooth,
leading to potentially correlated failure times within a patient.
These complexities extend to variable selection, a critical task in large cohort studies where the number of covariates is typically high.
While similar issues have been addressed under the proportional hazards (PH) model,
there is limited literature within the semiparametric multivariate accelerated failure time (AFT) framework.
To address these challenges, we propose an efficient weighted GEE approach combined with a
penalized variable selection procedure for stratified survival data under the AFT model.

Estimation methods for failure time data under stratified sampling designs have
predominantly focused on the weighted PH model framework
\citep{cox1972regression, kulich2004improving, samuelsen2007stratified, kang2009marginal, yan2017improving}.
However, the PH assumption may not hold in practice, and the hazard ratio is often
difficult to interpret in clinical settings \citep{wei1992accelerated}.
In contrast, the semiparametric AFT model offers a more interpretable alternative by directly relating
the logarithm of failure time to a linear combination of covariates,
providing effect estimates that are often more meaningful in clinical applications.
The semiparametric AFT model has been adapted to stratified sampling designs,
with case-cohort designs as a special case.
For example, \citet{nan2006censored}, \citet{kong2009case}, and \citet{chiou2015semiparametric}
extended the rank-based estimation method of \citet{tsiatis1990estimating} to accommodate such sampling schemes.
Additional developments by \citet{nan2009asymptotic} and \citet{yu2006hybrid} addressed more general settings
where covariates are systematically missing by design.
Nevertheless, these rank-based approaches do not fully account for within-cluster dependence,
which is particularly relevant in the retrospective dental study described earlier,
where failure times from the same patient may be correlated.
As an alternative to the rank-based approaches, \citet{yu2007buckley} and \citet{yu2011buckley} investigated the
Buckley–James (BJ) estimator \citep{buckley1979linear} within the AFT framework under case-cohort designs.
Compared to rank-based estimators, BJ estimators are more intuitively derived,
as they handle censored data through the least-squares principle.
\citet{lai1991large} rigorously established their asymptotic properties,
and \citet{ritov1990estimation} showed their asymptotic equivalence to the class of rank-based estimators
proposed when not adjusted for biased sampling. 
Despite these strengths, the use of BJ estimators has been limited by the lack of efficient and stable computational methods.
To address this, \citet{jin2006least} developed an iterative least-squares procedure that significantly improved the feasibility of BJ estimation.
Building on this development, we extend the BJ approach for semiparametric AFT models
to address the aforementioned challenges. 

Most existing analyses of stratified sampled data focus on univariate failure times,
even though clustered failure times are increasingly common in epidemiological studies.
The within-cluster dependence can be addressed in several ways. One common approach is the random effects model, which explicitly models the within-cluster correlation through subject-specific random effects. Alternatively, 
the generalized estimating equation (GEE) approach, which  originally developed by \citet{liang1986longitudinal} for uncensored clustered data, 
has become widely used due to its efficiency and robustness to misspecification of the correlation structure. The selection between these methods depends upon the goal of the analysis: 
the random effect model provides the subject-specific effects of covariates on the response, while the GEE methods focus on the population-average covariate effect. In this paper, we adopt the GEE framework, as our primary interest lies in estimating the effect of covariates on population-mean responses. \citet{chiou2014marginal} incorporated GEE techniques into the Buckley–James (BJ) framework
to account for within-cluster correlation in right-censored data.
We build on their work by extending the GEE-embedded BJ approach to
accommodate within-cluster dependence under a clustered stratified sampling design.

Although stratified sampling designs aim to reduce the cost and effort of covariate collection,
a large number of covariates may still be measured within the selected subcohort.
Identifying the covariates most strongly associated with the event of interest remains a key objective.
To address this, various penalized variable selection methods such as the least absolute shrinkage and selection operator (Lasso),
smoothly clipped absolute deviation (SCAD) \citep{fan2002variable}, adaptive Lasso (aLasso) \citep{zou2006adaptive},
and elastic net \citep{zou2005regularization} have been applied to the Buckley–James (BJ) estimator
within the AFT framework \citep{wang2008doubly, johnson2008penalized}.
However, existing penalized BJ approaches \citep{johnson2008variable, johnson2008penalized} can suffer from non-convergence issues,
as demonstrated by \citet{cai2009penalized},
due to instability in the underlying unpenalized BJ estimation procedure.
In addition, these variable selection methods are typically developed for full cohort data
and do not extend naturally to stratified sampling designs,
where appropriate adjustment for the sampling scheme is needed.
More recently, \citet{ni2016variable} proposed a variable selection procedure
for a case-cohort study using the SCAD penalty under the PH model.
However, to the best of our knowledge, penalized variable selection methods
for the AFT model under stratified sampling designs have not yet been formally investigated.

In this article, we propose a penalized weighted least-squares estimation method for analyzing clustered failure time data arising from stratified sampling designs.
The proposed approach addresses key challenges,
including (1) incomplete data due to partial cohort sampling, (2) computational instability often encountered in penalized BJ estimators, and (3) within-cluster dependence among failure times.
Inspired by the iterative least-squares algorithm of \citet{jin2006least},
our method employs an expectation-maximization (EM)-like procedure to improve computational stability,
while simultaneously incorporating GEE techniques to efficiently account for within-cluster correlation.
We establish the asymptotic properties of the resulting penalized weighted least-squares estimator.
Simulation studies highlight the importance of incorporating sampling weights to correct for bias
induced by stratified designs and
demonstrate the robustness of our method to misspecification of the within-cluster correlation structure.

This article is organized as follows. We present the weighted least-squares estimation for the AFT model
with clustered data from stratified sampling in Section \ref{sec:2}. In Section \ref{sec:3},
we detail the estimation method and establish the asymptotic properties of the proposed variable selection procedure. 
Section \ref{sec:4} examines practical considerations for implementation. 
In Section \ref{sec:5}, we evaluate the validity of the proposed variable selection procedure through simulation studies and apply the proposed method to the dental dataset in Section \ref{sec:6}. 
Lastly, Section \ref{sec:7} concludes with a discussion and future research directions.

\section{Buckley-James estimation under stratified sampling\label{sec:2}}

Consider a random sample of $n$ independent clusters with $K_i$ members in the $i$th cluster. 
For $i \in \{1, \ldots, n\}$ and $k \in  \{1, \ldots, K_i\}$,
let $T_{ik}$, $C_{ik}$, and $\bmX_{ik}$ be
the log-transformed failure time, the log-transformed censoring time, and the $1\times p$ vector of covariates, respectively.
We assume that the cluster size $K_i$ is assumed to be observed and
$T_{ik}$ and $C_{ik}$ are conditionally independent given $\bmX_{ik}$.
Define the observed failure time $Y_{ik} = \min(T_{ik}, C_{ik})$ and the censoring indicator 
$\Delta_{ik} = I(T_{ik} < C_{ik})$, where $I(\cdot)$ is the indicator function. 
We consider the following multivariate AFT model 
\begin{equation}
  \bmT_{i} = \bmX_{i} \bm{\beta}+\bm{\varepsilon}_{i}, \mbox{  ~} i=1,\dots,n,
  \label{eqn:aft}
\end{equation}
where $\bmT_{i}=(T_{i1}, \dots, T_{iK_i})^\top$, 
$\bmX_i=(\bmX_{i1}^\top, \dots, \bmX_{iK_i}^\top)^\top$,
$\bm{\beta}=(\beta_1,\dots,\beta_p)^\top$ is an unknown vector of regression coefficients,
and $\bm{\varepsilon}_{i} = (\varepsilon_{i1}, \dots, \varepsilon_{iK_i})^\top$ are 
unspecified random error vectors and are independent across clusters. 
We assume  $(\varepsilon_{i1},\dots,\varepsilon_{iK^\ast})^\top$ and $(\varepsilon_{j1},\dots,\varepsilon_{jK^\ast})^\top$ are identically distributed for 
$1\leq i \ne j\leq n$ and $K^\ast \leq \min(K_i,K_j)$, but
$\{\varepsilon_{i1},\dots,\varepsilon_{iK^\ast}\}$ may be correlated within a cluster.
For ease of discussion, we assume that $K_i=K$ and $\bm{\varepsilon}_i$ have the same marginal distribution $F$ for $i \in \{1,\dots,n\}$. 

If complete covariates information is available for the full cohort, regression coefficient $\bm{\beta}$ in \eqref{eqn:aft} can be estimated by the BJ estimator \citep{buckley1979linear} using the following least-squares estimating equation \eqref{bjscore}
\begin{equation}
    \sum_{i = 1}^n(\bmX_i - \bar \bmX)^\top
  \left\{\hat \bmY_i(\bm{\beta}) - \bmX_i \bm{\beta}\right\}=\bm{0}. 
\label{bjscore}
\end{equation}
where $\hat \bmY_i(\bm{\beta})=(\hat Y_{i1}(\bm{\beta}),\dots,\hat Y_{iK}(\bm{\beta}))^\top$
with $
\hat Y_{ik}(\bm{\beta}) = \Delta_{ik} Y_{ik}+ (1 - \Delta_{ik})\left[ \{\int_{e_{ik}(\bm{\beta})}^\infty u \,d \hat{F}_{n}^{\bm{\beta}}(u)\}/[1 - \hat{F}_{n}^{\bm{\beta}}\{e_{ik}(\bm{\beta})\}] +\bmX_{ik} \bm{\beta}\right]
$. 
In this estimating equation, $\hat Y_{ik}(\bm{\beta})$ replaced the censored $T_{ik}$ by an estimation of the conditional expectation $\mathrm{E}_{\bm{\beta}}(T_{ik}|Y_{ik}, \Delta_{ik}, \bmX_{ik})$ evaluated at regression coefficient $\bm{\beta}$, where the conditional distribution is estimated using Kaplan-Meier estimator from pooled residuals $e_{ik}(\bm{\beta}) = Y_{ik} - \bm{X}\bm{\beta}$, denoted by $\hat{F}_n^{\bm{\beta}}$, for the error distribution $F$.

Estimating equation \eqref{bjscore} cannot be directly applied under the stratified sampling design where the covariates are only observed in a selected subcohort. 
Using complete case analysis to evaluate \eqref{bjscore} could result in biased inference because the stratified sample is a biased population representation. 
To address this, the equation \eqref{bjscore} must be adjusted with appropriate weights to account for the stratified random sampling scheme. 
Suppose the full cohort is divided into $S$ mutually exclusive strata, with a stratified sample generated by randomly drawing from each stratum with different inclusion probabilities $p_{n,s}=\tilde{n}_s/n_s$, $s \in \{1,\dots,S\}$, where $\tilde{n}_s$ and $n_s$ are the numbers of sampled clusters and cluster in the $s$th stratum, respectively. 
Let $\psi_{is}$ denote the stratum indicator, where $\psi_{is}=1$ if the $i$th cluster was in the $s$th stratum, and $\psi_{is}=0$ otherwise. 
Similarly, let $\xi_i$ denote the sampling indicator, where $\xi_i=1$ if the $i$th cluster is selected and $\xi_i=0$ if not. 
 One appropriate sampling weight for  the $i$th cluster in the stratified sample is the inverse of the inclusion probability, given by $\omega_i=\sum_{s=1}^S \xi_i \psi_{is}/\sum_{s=1}^S \psi_{is}p_{n,s}$.  
Correspondingly, the weight-adjusted imputed response for $k$th subject in $i$th cluster is
\begin{equation}
\hat Y_{ik,\omega}(\bm{\beta}) = \Delta_{ik} Y_{ik}+ (1 - \Delta_{ik})\left[ \frac{\int_{e_{ik,\omega}(\bm{\beta})}^\infty u \,\dif \hat{F}_{n,\omega}^{\bm{\beta}}(u)}{1 - \hat{F}_{n,\omega}^{\bm{\beta}}\{e_{ik,\omega}(\bm{\beta})\}} +\bmX_{ik} \bm{\beta}\right], \mbox{ for } i=1,\dots,n, ~\mbox{ and }~ k=1,\dots,K, 
\label{eqn:imputeY}
\end{equation}
where 
\begin{equation}
\hat{F}_{n,\omega}^{\bm{\beta}}(t) = 1 - \prod_{i,k: e_{ik,\omega}(\bm{\beta}) < t}\left[1 - \frac{\omega_{i}\Delta_{ik}}
{\sum_{j = 1}^n\sum_{l=1}^{K}\omega_{j}I\{e_{jl,\omega}(\bm{\beta})\ge e_{ik,\omega}(\bm{\beta})\}}\right]. 
\label{km}
\end{equation}
The weighted and pooled Kaplan-Meier estimator $\hat{F}_{n, \omega}^{\bm\beta}(t)$  
is shown as a consistent estimator of the identical marginal distribution of errors in Lemma 1 in the Appendix. 

We incorporate the GEE method into the estimating procedure to improve estimation efficiency while addressing within-cluster correlation.
As a result,
the weighted estimating equation for clustered failure time data from the stratified sampling is given by  
\begin{equation}
\bm{U}_{n,\omega}(\bm{\beta},\bm{\alpha}) = \sum_{i = 1}^n\omega_i(\bmX_i - \bar\bmX_\omega)^\top\bm{\Omega}^{-1}\{\bm{\alpha}\}\left\{\hat\bmY_{i, \omega}(\bm{\beta}) - \bmX_i\bm{\beta}\right\}=\bm{0},
\label{wgee1}
\end{equation}
where 
$\bar\bmX_\omega=(\sum_{i = 1}^n \omega_i\bmX_i )/(\sum_{i = 1}^{n}\omega_{i})$ is the weighted mean of covariates, 
$\hat\bmY_{i,\omega}(\bm{\beta}) = \{\hat{Y}_{i1,\omega}(\bm{\beta}),\ldots, \hat{Y}_{iK,\omega}(\bm{\beta})\}^\top$ is the vector of the weight-adjusted imputed response for stratified samples, and $\bm{\Omega}\{\cdot\}$ is an $K\times K$ 
working covariance matrix with a finite-dimensional correlation parameter vector $\bm{\alpha}$. 
When the working covariance structure is specified, $\bm{\alpha}$ can be estimated by methods of the moment \citep{liang1986longitudinal} from the weighted version of the residuals $e_{ik,\omega}$. 
Although the estimator of $\bm{\alpha}$, denoted by $\hat{\bm{\alpha}}$, is involved in the iteration, we treat it as a nuisance variable in our estimation procedure. 
The consistency and asymptotically normality of the estimates inherit from the initial value and do not depend on the correct specification of the correlation structure $\bm{\Omega}$.  
However, the correct specification of $\bm{\Omega}$ can improve the efficiency of the parameter estimation. 

\section{Variable selection with the penalized estimating equation\label{sec:3}}

\subsection{Penalized estimating equation}

The penalized estimating equation under stratified sampling is defined as 
\begin{equation}
\tilde{\bm{U}}_{n,\omega}(\bm{\beta},\bm{\alpha}) = \bm{U}_{n,\omega}(\bm{\beta}, \bm{ \alpha}) - n \mathbf{q}_{\lambda_n} (|\bm{\beta}|)=\bm{0},
\label{penal0}
\end{equation}
where 
$\mathbf{q}_{\lambda_n}(|\bm{\beta}|)= \{p^\prime_{\lambda_n}(|{\beta_1|})\cdot \sgn(\beta_1),\dots,p^\prime_{\lambda_n}(|{\beta_p|})\cdot \sgn(\beta_p)\}^\top$,
$p^\prime_{\lambda_n}(\cdot)$ is the first derivative of the penalty function $p_{\lambda_n}(\cdot)$ 
with a nonnegative tuning parameter $\lambda_n$, and $\sgn(\cdot)$ is the sign function.
A proper penalty function shrinks the coefficients of less influential covariates toward zero
while retaining the more relevant covariates in the final model.
Among commonly used penalties, the smoothly clipped absolute deviation (SCAD) penalty satisfies the oracle property \citep{fan2001variable},
meaning that the resulting penalized estimator is asymptotically equivalent to the estimator obtained as if the true underlying model were known.
In this paper, we illustrate the proposed method with the SCAD penalty function for variable selection
due to its favorable theoretical properties and its ability to balance sparsity and unbiased estimation of large coefficients.
However, the proposed method is flexible and can be readily extended to incorporate other penalty functions.
The SCAD penalty function \citep{fan2001variable} is given by
\begin{equation}
    p^\prime_{\lambda_n}(|\beta|)= \lambda_n I(|\beta|< \lambda_n) + \frac{(a\lambda_n-|\beta|)_+}{(a-1)}I(|\beta|\geq \lambda_n),
\end{equation}
where $a >0$ and $(c)_+=\max\{0, c\}$ for $c\in \mathbb{R}$. 
In simulation and data analysis, we considered $a = 3.7$ in the SCAD penalty as suggested in \citet{fan2001variable}.

\subsection{Estimation methods}
Solving \eqref{penal0} for $\bm\beta$ is challenging because $\hat{Y}_{ik,\omega}(\bm{\beta})$ involved
in the BJ estimation in \eqref{bjscore} is a discontinuous and piecewise linear function of $\bm{\beta}$.
This non-smooth structure can lead to instability in the optimization process,
causing the solution to oscillate between multiple candidate estimates and hindering convergence \citep{cai2009penalized}.
To address these convergence challenges, \citet{jin2006least} proposed a reliable iterative procedure that reduces the problem to solving a sequence of linear estimating equations, each with a closed-form solution.
\citet{chiou2014marginal} extended the procedure to account for within-cluster correlation under the GEE framework. However, this approach is not directly applicable to variable selection problems, as closed-form solutions for penalized $\bm{\beta}$ estimators are generally unavailable due to the non-differentiable nature of penalty terms at the origin. To enable regularization, we extend the iterative procedure of \citet{jin2006least} by incorporating the algorithm for penalized GEE from \citet{wang2012penalized}, which employs the minorization-maximization approach to handle the penalty \cite{hunter2005variable} and the Newton-Raphson method for solving the GEE.

The proposed estimation method begins by fixing $\hat{\bm{Y}}_{i,\omega}(\bm{b})$ at an initial estimator $\bm{b}$ and  reformulating the estimating function $\bm{U}_{n,\omega}(\bm{\beta}, {\bm{\alpha}})$ to a linear function of $\bm{\beta}$.
Specifically, given $\bm{b}$, the penalized GEE is given by
\begin{equation}
    \tilde{\bm{U}}_{n,\omega}(\bm{\beta},
\bm{b}, \bm{\hat{\alpha}}(\bm{b})) = \sum_{i = 1}^n(\bmX_i - \bar\bmX_\omega)^\top\omega_i\bm{\Omega}^{-1}(\bm{\hat{\alpha}}(\bm{b}))\left\{\hat\bmY_{i, \omega}(\bm{b}) - \bmX_i\bm{\beta}\right\}- n \mathbf{q}_{\lambda_n} (|\bm{\beta}|)=\bm{0}.
\label{penalty1}
\end{equation} 
This reformulation facilitates the use of the penalized GEE algorithm to obtain a penalized estimator $\hat{\bm{\beta}}$ of $\bm{\beta}$ based on $\bm{b}$.
The updated $\hat{\bm{\beta}}$ is then used to refine $\bm{b}$ in $\hat{\bm{Y}}_{i,\omega}(\bm{b})$. 
Repeating this process yields a two-layer iterative algorithm for solving \eqref{penal0}: 
 the inner layer solves the penalized GEE for a fixed $\bm b$, while the outer layer iteratively updates 
$\bm b$ until convergence.

In particular, let $s$ denote the iteration step within the inner layer, and $\nu$ represent the outer layer index.  A Newton-Raphson recursive formula can be utilized to solve the GEE \eqref{penalty1} with respect to $\bm{\beta}$. 
Given $\bm{b}$ and $\lambda_n\geq0$,  set $\hat{\bm{\beta}}^{( 0)}_\omega =\bm{b}$, we update $\hat{\bm{\beta}}_\omega^{(s+1)}$ by 
\begin{equation}
\begin{split}
\hat{\bm{\beta}}_\omega^{( s+1)} =&~ \hat{\bm{\beta}}_\omega^{( s)}  + \left \{\bm{H}_{n,\omega}(\bm{\hat{\alpha}}(\bm{b}))  + n\bm{G}_{n,\omega}(\hat{\bm{\beta}}^{( s)}_\omega)  \right\}^{-1}  \times \left\{ \bm{U}_{n,\omega}\{\hat{\bm{\beta}}_\omega^{( s)}, \bm b, \bm{\hat{\alpha}}(\bm{b})\}  - n\bm{G}_{n,\omega}(\hat{\bm{\beta}}^{( s)}_\omega) \hat{\bm{\beta}}^{( s)}_\omega \right\},
\end{split}
\label{updatetheta}
\end{equation}
where 
\begin{equation}
    \bm{H}_{n,\omega}(\bm{\hat{\alpha}}(\bm{b}))
    = \sum_{i=1}^n \bmX_i^\top \omega_i \bm{\Omega}^{-1}(\bm{\hat{\alpha}}(\bm{b}))(\bmX_i - \bar{\bmX}_\omega),
\end{equation}

\begin{equation}
    \bm{U}_{n,\omega}(\hat{\bm{\beta}}^{( s)}_\omega, \bm{b}, {\bm{\hat{\alpha}}(\bm{b})}) = \sum_{i = 1}^n(\bmX_i - \bar\bmX_\omega)^\top\omega_i\bm{\Omega}^{-1}(\bm{\hat{\alpha}}(\bm{b}))\left(\hat\bmY_{i, \omega}(\bm{b}) - \bmX_i\hat{\bm{\beta}}^{(s)}_\omega\right),
\end{equation}  
\begin{equation}
    \bm{G}_{n,\omega}(\hat{\bm{\beta}}^{( s)}_\omega)
    =~ \mathrm{diag}\left\{\frac{p^\prime_{\lambda_n}(|{\hat{\beta}_{1,\omega}^{( s)}|)}}{\zeta+|\hat{\beta}_{1,\omega}^{( s)}|},\cdots,\frac{p^\prime_{\lambda_n}(|{\hat{\beta}_{p,\omega}^{( s)}|)}}{\zeta+|\hat{\beta}_{p,\omega}^{( s)}|} \right\}, \mbox{ for } \lambda>0. \notag
\end{equation}
The matrix $\bm{G}_{n,\omega}(\hat{\bm{\beta}}^{( s)}_\omega)$  reduces to a zero matrix when $\lambda_n=0$,
and $\zeta$ serves as a small positive numerical adjustment to handle zero elements in $\hat{\bm{\beta}}$
during iterations. 
The correlation parameter $\bm{\alpha}$ is estimated by $\bm{\hat{\alpha}}(\bm{b})$, as described in Section \ref{sec:4}.
In both the simulation study and data analysis,
we set $\zeta=10^{-6}$ and use the convergence criterion $\|\hat{{\bm{\beta}}}_\omega^{(s+1)} - \hat{{\bm{\beta}}}_\omega^{(s)}\|_{\infty} \leq \gamma$
for updating equation \eqref{updatetheta}.

At convergence of the Newton-Raphson update in equation \eqref{updatetheta}, the estimate $\hat{\bm{\beta}}_\omega^{(s)}$ is a solution to GEE \eqref{penalty1} for a fixed $\bm b$.
Once convergence is achieved in the inner layer,
the outer-layer estimate $\bm b$ is updated using the resulting $\hat{\bm{\beta}}_\omega^{(s)}$. 
The inner-layer iterations are then reinitialized by setting $s = 0$, and the process is repeated.
The algorithm continues until the outer-layer sequence 
$\bm{b}^{(\nu)}$ satisfies the convergence criterion $\|{\bm{b}}^{(\nu+1)} - {\bm{b}}^{(\nu)}\|_{\infty} \leq \gamma $,
at which point convergence in $\nu$ is declared.
Simulation results show that the number of iterations required for the outer layer is influenced by the censoring rate.
Lower censoring generally leads to fewer iterations, as less imputation is needed.
For example, under a 50\% censoring rate with standard normal errors,
the outer layer required
an average of $35\%$ less iterations than a 90\% censoring rate under our simulation settings. 
In contrast, the inner layer typically converges within 2 to 5 iterations and is relatively insensitive to the censoring level, since no imputation is required.
In all simulations, the convergence threshold was set to $\gamma = 10^{-3}$.
The complete procedure is outlined in Algorithm 1.

\begin{algorithm}[ht]
\caption{Algorithm for solving $\tilde{\bm{U}}_{n,\omega}(\bm{\beta},\bm{\alpha}) = \bm{0}$ for a fixed $\lambda_n\geq0$.}
\label{alg:paftgee}
Obtain initial estimates ${\bm{b}}$ and set $\hat{{\bm{\beta}}}^{(0)}_\omega = \bm{b}$;\\
Select the best tuning parameter $\lambda_n\geq0$;\\
Initialize  $\nu = 0$ and set  $\bm{b}^{(0)} = \bm{b}$;\\
 \Repeat{ $\bm{b}$ converged in $\nu$, i.e., $\|{\bm{b}}^{(\nu+1)} - {\bm{b}}^{(\nu)}\|_{\infty} \leq \gamma $  }{
 
	Calculate $\hat{F}_{n,\omega}^{\bm{b}}(\cdot)$ and $\hat Y_{ik,\omega}(\bm{b})$;\\
	 Initialize  $s = 0$; \\
		\Repeat{$\hat{\bm{\beta}}^{(s+1)}_\omega$ converged in $s$, i.e., $\|\hat{{\bm{\beta}}}_\omega^{(s+1)} - \hat{{\bm{\beta}}}_\omega^{(s)}\|_{\infty} \leq \gamma $}{
    \noindent 
   Calculate   $\hat{\bm{\alpha}}=\hat{\bm{\alpha}}(\hat{\bm{\beta}}_\omega^{(s)})$,  $\bm{H}_{n,\omega}(\bm{\hat{\alpha}}(\bm{b}))$
			and 
$\bm{G}_{n,\omega}(\hat{\bm{\beta}}^{(s)}_\omega)
    = \mathrm{diag}\left\{\frac{p^\prime_{\lambda_n}(|{\hat{\beta}_{1,\omega}^{(s)}|)}}{\zeta+|\hat{\beta}_{1,\omega}^{(s)}|},\cdots,\frac{p^\prime_{\lambda_n}(|{\hat{\beta}_{p,\omega}^{(s)}|)}}{\zeta+|\hat{\beta}_{p,\omega}^{(s)}|} \right\};$\\
Update $\hat{\bm{\beta}}_\omega^{(s+1)} = \hat{\bm{\beta}}_\omega^{(s)}  + \left [\bm{H}_{n,\omega}(\bm{\hat{\alpha}}(\bm{b}))  + n\bm{G}_{n,\omega}(\hat{\bm{\beta}}^{(s)}_\omega)  \right]^{-1}
\times \left[ \bm{U}_{n,\omega}(\hat{\bm{\beta}}_\omega^{(s)}, \bm b, (\bm{\hat{\alpha}}(\bm{b}))  - n\bm{G}_{n,\omega}(\hat{\bm{\beta}}^{(s)}_\omega) \hat{\bm{\beta}}^{(s)}_\omega \right];$\\
Set $s \leftarrow s + 1$;\\
		}
	 Update $\bm{b} = \hat{{\bm{\beta}}}_\omega^{(s+1)}$ and set  $\bm{b}^{(\nu+1)} = \bm{b}$;\\
     $\nu \leftarrow \nu + 1$;\\
	}
return $\bm{b}^{(\nu+1)}$. 

\end{algorithm}

The traditional $M$-fold CV procedure, which randomly divides the entire data into $M$
approximately equal-sized non-overlapping subsets, may not be suitable for selecting the tuning parameter
$\lambda_n$ in stratified sampling designs.
This is because random partitioning can produce unbalanced subsets,
where the distribution of strata or case-control status is not preserved across the $M$ folds. 
For example, in stratified case-cohort designs with heavy censoring,
traditional $M$-fold CV could result in a training set with no cases,
leading to biased training parameters and prediction errors. 
To mitigate this issue, we modify the CV procedure to preserve the stratified structure of the original data.
Specifically, we apply the standard CV split within each stratum and then aggregate the corresponding
$m$-th folds across strata to construct the final $m$-th CV sample.
This stratified approach ensures that each fold retains a similar proportion of cases and controls as in the full dataset,
thereby reducing training bias and improving the stability of prediction error estimates.
For a fixed $\lambda_n$, we remove one fold and obtain 
the estimator of $\bm{\beta}$, denoted by $\hat{\bm{\beta}}_\omega(\lambda_n)$,  from the remaining folds. 
The predicted error for the $j$th subject in $i$th cluster
in the removed fold is then given by
$
    \PE_{ij}(\lambda_n)=\{\hat{Y}_{ij,\omega}(\hat{\bm{\beta}}_{\omega})-\bm{X}_{ij}\hat{\bm{\beta}}_\omega(\lambda_n)\}^2
$, where $\hat{Y}_{ij,\omega}(\hat{\bm{\beta}}_{\omega})$ is the weight-adjusted imputed response and $\hat{\bm{\beta}}_{\omega}$ is the unpenalized estimator obtained from the removed fold.
    By averaging the predicted errors across all observations for all $M$ possible combinations of training and validation subsets,
    the overall prediction performance for a given $\lambda_n$ is measured by the weighted average
$
\mu_{\omega}(\lambda_n) 
= \sum_{i=1}^n\omega_i\sum_{j=1}^{K} \PE_{ij}(\lambda_n)/\sum_{i=1}^n \omega_{i}{K}.
$

The optimal $\lambda_n$ is typically selected as the one that results in the smallest mean
prediction error among a fine grid of $\lambda_n$ values, denoted by
$\lambda_n^{ \mbox{CV}} = \arg\min_\lambda\mu_\omega(\lambda_n)$.
An alternative method for selecting the optimal $\lambda_n$ is the one-standard error rule 
\citep{friedman2001elements}, which favors a more parsimonious model by choosing the largest
$\lambda_n$ whose prediction error is within one standard error of $\lambda_n^{ \mbox{CV}}$.
Let $\lambdaSE$ be the $\lambda_n$ selected by the one-standard error rule, then
$\lambdaSE$ is the largest $\lambda_n$ that satisfies 
$\mu_{\omega}(\lambda_n) \leq \mu_{\omega}(\lambdaCV)+\mathrm{SE}(\lambdaCV),$ where 
\begin{equation}
\mathrm{SE}(\lambdaCV) = \sqrt{\frac{\sum_{i=1}^n \omega_i \sum_{j=1}^{K}\{PE_{ij}(\lambdaCV)-\mu_{\omega}(\lambdaCV)\}^2}{\sum_{i=1}^n \omega_i(n-1)}}.\notag
\end{equation}

\subsection{Asymptotic properties}

The asymptotic theory provided by \citet{johnson2008penalized} for the penalized BJ estimator is not applicable to our framework for several reasons. As discussed in \citet{wang2014variable}, the penalized estimator for stratified sampling designs require further investigation both in theory and in practice. In addition, the assumption of independent failure times, which is essential in \citet{lai1991large} and \citet{johnson2008penalized}, is not satisfied in the present of correlated failure times within the GEE framework. Moreover, our iterative inference procedure for penalized estimators incorporates a linearization step of the BJ estimating function that is not present in existing work. Therefore, we study the asymptotic properties for the proposed estimators in the following theorems.

To simplify notation, we let $\|\cdot\|$ denote the Euclidean norm when applied to vectors and 
the Frobenius norm when applied to matrices.
We also use $O_p(\cdot)$ and $o_p(\cdot)$ to denote probability boundedness, 
while $O(\cdot)$ and $o(\cdot)$ refer to almost-sure boundedness.
We impose the following conditions:

\begin{description}
    \item[C1]  $||\bmX_{i}||\leq C_0$ for all $i$ and some nonrandom constant $C_0$.  
    \item[C2] The marginal distribution function $F$ of error vector $\bm{\varepsilon}_i$ has a finite second moment. 
The density function of $F$ exists, denoted by $f$, is twice differentiable such that $\int^\infty_{-\infty} \{f^\prime(t)/f(t)\}^2 \, dF(t)$ and $\int_{-\infty}^{\infty} \sup_{|h| \leq \delta} \{|f^\prime(t+h)|+|f''(t+h)|\} \, dt$ are bounded for some $\delta>0$, where $f'$ and $f''$ are the first and second derivatives of $f$. 
    \item[C3] For some $\theta > 0$ and for every $i = 1, \ldots, n$, 
    $\sup_{k\in \{1,\dots,K\}}\mathbb{E}\{\exp(\theta \varepsilon_{ik}^-)\} + \mathbb{E}\{\exp(\theta C_{ik}^-)\} < \infty$, where $a^-$ is the negative part of any value $a$, that is, $a^-=|a|I_{\{a \leq 0\}}$ and $I(\cdot)$ is the indicator function. 
    \item[C4] When $h \to 0$ and $nh \to \infty$, 
$\sup_{\|\bm{\beta}\|<B, |t|< \infty}
\sum_{i=1}^n 
\sum_{j=1}^K \Pr(t<C_{ij}-\bm{X}_{ij}\bm{\beta} \leq t+h)=O(nh)$ for some $B>0$.
    \item[C5] There is a constant $a$ such that $\Pr(Y_{ij} - \bm{X}_{ij}\bm{\beta} \geq a) \geq \kappa_0 > 0$ for all $\bm{\beta}$ satisfying $\|\bm{\beta}\|<B$. 
    \item[C6] The sampling weights $\omega_i, i=1,\dots,n,$ defined in Section 2.2 do not depend on any parameter and satisfy the boundedness condition $h_1 \leq \omega_i \leq h_2$ for some positive and finite constants 
    $h_1$ and $h_2$. 
    \item[C7] The initial estimator $\bm{b}$ of true $\bm{\beta}_0$ is consistent and $\sqrt{n}(\bm{b}-\bm{\beta}_0)$ is asymptotically normal.
    \item[C8] The true covariance matrix $\bm{\Omega}_0$ of $\bm{T}_i=(T_{i1},\dots,T_{iK})^\top$ 
has positive and bounded eigenvalues. 
The estimated working covariance matrix $\bm{\Omega}(\bm{\hat{\alpha}}(\bm{b}))$ satisfies $||\bm{\Omega}^{-1}(\bm{\hat{\alpha}}(\bm{b})) - \bar{\bm{\Omega}}^{-1}||=O_p(n^{-1/2})$, where $\bar{\bm{\Omega}}$
is a constant positive definite matrix and not necessarily to be $\bm{\Omega}_0$. 
    \item[C9] For every $t$ such that $F(t)<1$, as $n\to \infty$, assume that $\lim_{n\to \infty} n^{-1}\sum_{i = 1}^n \sum_{j=1}^K \Pr(C_{ij}-\bmX_{ij}\bm{\beta}_0\geq t) \to \Gamma_{0}(t)$, $\lim_{n\to \infty} n^{-1}\sum_{i = 1}^n \sum_{j=1}^K \bmX_{ij}^\top \Pr(C_{ij}-\bmX_{ij}\bm{\beta}_0\geq t) \to \bm{\Gamma}_{1}(t)$, $\lim_{n\to \infty}  n^{-1}\sum_{i = 1}^n \sum_{j=1}^K  \omega_i  [(\bmX_i - \bar\bmX_\omega)^\top\bm{\Omega}^{-1}\{\bm{\alpha}(\bm{\beta})\}]_{lj}\Pr(C_{ij}-\bmX_{ij}\bm{\beta}_0\geq t)\to \Gamma^{\omega,\ell}_1(t)$, and $ n^{-1}\sum_{i = 1}^n \sum_{j=1}^K  \omega_i \bmX_{ij}^\top [(\bmX_i - \bar\bmX_\omega)^\top\bm{\Omega}^{-1}\{\bm{\alpha}(\bm{\beta})\}]_{lj}\Pr(C_{ij}-\bmX_{ij}\bm{\beta}_0\geq t) \to \bm{\Gamma}^{\omega,\ell}_2(t)$. Here, $[\cdot]_{lj}$ represents the element at $l$th row and $j$th column for a given matrix. 
\end{description}


Conditions C1 -- C4 are similar to those outlined in \citet{lai1991large} for establishing the asymptotic properties of the Buckley-James estimators. 
Condition C5 is common 
in survival analysis to ensure the tail stability of Kaplan-Meier estimates, such as \citet{yu2011buckley} and \citet{yang1997generalization}.
For condition C6, it is natural to assume that sampling weights $\omega_i$ are finite and not a random variable. Condition C7 is necessary for deriving the consistency and asymptotic normality of the Buckley-James estimators under iterative estimation approaches, which is also required in \citet{jin2006least} and \citet{chiou2014marginal}. 
Condition C8 is a standard assumption of the GEE for multivariate data as \citet{wang2012penalized}.
Condition C9 is fundamental for deriving the asymptotic linearity of 
the estimating function and the consistency of Buckley-James estimators, which parallels the assumptions in \citet{lai1991large} and \citet{jin2006least}.


Suppose the true regression parameter $\bm\beta_0$ under~\eqref{eqn:aft} 
is partitioned into two parts:
$\bm\beta_0 = (\bm\beta_{10}^\top, \bm\beta_{20}^\top)^\top$,
where $\bm\beta_{10}$ is a $d\times1$ nonzero vector and $\bm\beta_{20}$ is a $(p - d)\times1$ vector of zeros.
The exact solution of $\tilde{\bm{U}}_{n,\omega}(\bm{\beta},\bm{b},\bm{\alpha})=\bm{0}$ with respect to $\bm{\beta}$ may not exist, as the penalty term $\bm{q}_\lambda(|\bm{\beta}|)$ is not continuous in $\bm\beta$. 
Given an initial estimator $\bm b$ that is consistent and asymptotically normal for $\bm{\beta}_0$, 
we first establish the existence of a penalized weighted GEE estimator obtained from \eqref{penalty1}, 
and then prove that it possesses the oracle property. 
In particular, we show that the variable selection procedure yields a penalized estimator that is 
asymptotically equivalent to the oracle estimator derived under the assumption that the true model is known. 
Theorems 1 and 2 formally state these results, with proofs provided in the Appendix. 
A tail modification is applied to the estimating functions to stabilize the Kaplan–Meier estimator, following the approach of \citet{lai1991large}.

\begin{theorem}
Assume that the tuning parameter $\lambda_n$ satisfies $\lambda_n \to 0$ and $\sqrt{n}\lambda_n \to \infty$ as $n \to \infty$. 
Under conditions (C1) -- (C9), if $\max\{p^{\prime \prime}_{\lambda_n}(|\beta_{j0}|):\beta_{j0} \neq 0\} \to 0$ as $n\to\infty$, then given a $\sqrt{n}$-consistent $\bm{b}$, there exists an approximated solution $\hat{\bm{\beta}}_\omega$ of $\tilde{\bm{U}}_{n,\omega}(\bm{\beta}, \bm{b}, \bm{\alpha})=\bm{0}$ such that $||\hat{\bm{\beta}}_\omega-\bm{\beta}_{0}||=O_p(a_n)$, where $a_n = n^{-1/2} + \max\{p^\prime_{\lambda_n}(|\beta_{j0}|):\beta_{j0} \neq 0 \}$.
 \label{theorem1}
\end{theorem}


Theorem \ref{theorem1} implies that, at each outer-layer iteration, 
a penalized weighted GEE estimator $\hat{\bm{\beta}}_\omega$ exists and is $\sqrt{n}$-consistent, 
provided that the tuning parameter $\lambda_n$ is chosen such that $a_n=O(n^{-1/2})$. 
Moreover, one can show that $\hat{\bm{\beta}}_\omega$ in Theorem \ref{theorem1} is also an approximate zero crossing 
of the estimating equation $\tilde{\bm{U}}_{n,\omega}(\bm{\beta}, \bm{b}, \bm{\alpha})=\bm{0}$, 
consistent with the definition of an approximate solution to penalized BJ estimating equation proposed by \citet{johnson2008penalized}. 
Specifically, $\hat{\bm{\beta}}_\omega$ satisfies 
$$\lim_{n\to \infty}\lim_{c\to 0+} n^{-1}\tilde{U}_{nj,\omega}(\hat{\bm{\beta}}_\omega+c \bm{u}_j, \bm{b}, \bm{\alpha})\tilde{U}_{nj,\omega}(\hat{\bm{\beta}}_\omega-c \bm{u}_j, \bm{b}, \bm{\alpha})\leq 0, 
\mbox{for } j = 1, \ldots, p,$$
where $\bm{u}_j$ is the $j$th canonical unit vector, and $\tilde{U}_{nj,\omega}(\cdot)$ 
is the $j$th element of $\tilde{\bm{U}}_{n,\omega}(\cdot)$.
In other words, $\tilde{U}_{nj,\omega}(\hat{\bm{\beta}}_\omega, \bm{b}, \bm{\alpha})$ changes its sign with a small perturbation of $j$th component of $\hat{\bm{\beta}}_\omega$, for $j \in \{d+1,\dots,p\}$, 
which characterizes it as an approximate zero crossing.

To establish the asymptotic properties of the proposed estimator, 
we first define $\bm{D}_\omega=\lim_{n \to \infty} n^{-1}\sum_{i = 1}^n(\bmX_i - \bar\bmX_\omega)^\top\omega_i\bm{\Omega}^{-1}\{\bm{\hat{\alpha}}\}\bmX_i$, 
which serves as the weighted analogue of the slope matrix in the least-squares estimating equation for uncensored 
data under the GEE framework. 
The existence of $\bm{D}_\omega$ is established in Theorem 2, whose proof is provided in the Appendix. 
In addition, we show the asymptotic linearity of the unpenalized estimating function 
$\bm{U}_{n,\omega}(\bm{\beta}, \bm{\alpha})$, and define its associated slop matrix as $\bm{A}_\omega$.
The $l$th row vector of $\bm{A}_\omega$ is given by  
\begin{equation}
    \bm{A}_{\omega,\ell}^{\top} = \int_{-\infty}^{\infty} \left[ \left\{\bm{\Gamma}^{\omega,\ell}_2(t)-\frac{\bm{\Gamma}_1(t)\Gamma^{\omega,\ell}_1(t)}{\Gamma_0(t)}\right\} \int_t^\infty \left\{1-F(s)\right\}\, \dif s \right]\,\dif \lambda(t), \mbox{ } l \in \{1,\dots,p\},
\end{equation}
where $\lambda(t)$ is the hazard function of the marginal distribution $F(t)$, $\Gamma_0(t)$ is a nonrandom function, $\bm{\Gamma}_1(t)$ is a $p$-dimensional vector of nonrandom function, $\Gamma_1^{\omega,\ell}(t)$ is a weighted nonrandom function, and $\bm{\Gamma}^{\omega,\ell}_2(t)$ is a $p$-dimensional vector of weighted nonrandom functions, as defined in condition (C9).  
We further define $\bm{B}_{\omega}$ such that its $l$th row vector is given by 
\begin{equation}
    \bm{B}_{\omega,\ell}^\top=\int_{-\infty}^{\infty} \left [ \left\{\bm{\Gamma}^{\omega,\ell}_2(t)-\frac{\bm{\Gamma}_1(t)\Gamma^{\omega,\ell}_1(t)}{\bm{\Gamma}_0(t)}\right\} \frac{(\int_t^\infty \{1-F(s)\}\, \dif s)^2}{1-F(t)} \right]\ \dif\lambda(t).
\end{equation}

\begin{theorem}
Under conditions (C1) -- (C9), given a $\sqrt{n}$-consistent and asymptotically normal estimator $\bm{b}$, the approximate solution $\hat{\bm{\beta}}_\omega=(\hat{\bm{\beta}}_{1,\omega},\hat{\bm{\beta}}_{2,\omega})^\top$ of $\tilde{\bm{U}}_{n,\omega}(\bm{\beta}, \bm{b}, \bm{\alpha})= \bm{0}$ satisfies the following asymptotic properties. As $n\to\infty$, \\ 
(1) with probability tending to 1, $\lim_{n \to \infty}\Pr(\hat{\bm{\beta}}_{2,\omega}=\bm{0})=1$;\\
(2) with probability tending to 1, 
$\tilde{U}_{nj,\omega}(\hat{\bm{\beta}}_\omega,
\bm{b},\bm{\alpha})=0, \mbox{ for } j \in \{1,\dots,d\}, \mbox{ and } ~ |n^{-1}\tilde{U}_{nj,\omega}(\hat{\bm{\beta}}_\omega,
\bm{b},\bm{\alpha})|\leq 
\lambda_n, \mbox{ for } j \in \{d+1,\dots,p\}$;\\
(3) define $\bm{\Lambda}_{11}=diag\{p^{\prime\prime}_{\lambda_n}(|\beta_{10}|), \dots, p^{\prime\prime}_{\lambda_n}(|\beta_{d0}|)\}$, $\bm{q}_{01}=\{p^\prime_{\lambda_n}(|{\beta_{10}|})\cdot \sgn(\beta_{10}),\dots,p^\prime_{\lambda_n}(|{\beta_{d0}|})\cdot \sgn(\beta_{d0})\}^\top$,
    \begin{equation}
        \sqrt{n}(\bm{A}_{11}^\omega-\bm{\Lambda}_{11})\left\{\hat{\bm{\beta}}_{1,\omega}-\bm{\beta}_{10}-(\bm{A}_{11}^\omega-\bm{\Lambda}_{11})^{-1}\bm{q}_{01}  \right\} + \sqrt{n}(\bm{A}_{11}^\omega-\bm{D}_{11}^\omega)(\bm{b}_1-\hat{\bm{\beta}}_{1,\omega})  \to_d N(\bm{0}, \bm{B}^\omega_{11}),
        \label{normal}
    \end{equation}
    where $\bm{b}_1$ is the first $d$ elements of $\bm{b}$,   $p^{\prime\prime}_{\lambda_n}(t)$ is the second derivative of penalty function $p_{\lambda_n}(t)$ for $t>0$, $\bm{A}^\omega_{11}$, $\bm{B}^\omega_{11}$, and $\bm{D}^\omega_{11}$ are the first $d\times d$ sub-matrix of nonrandom matrices $\bm{A}_\omega$, $\bm{B}_\omega$, and $\bm{D}_\omega$,  respectively.
\label{th2}
\end{theorem}

Properties $(1)$ and $(3)$ of Theorem \ref{th2} together provide the oracle property of 
the penalized weighted estimator $\bm{\hat{\beta}}_\omega$ at each outer-layer iteration, 
assuming an appropriate penalty function such as SCAD. 
Property $(2)$ describes more precisely the approximate solution of penalized weighted GEE \eqref{penalty1}. 
As the censoring rate approaches zero, $\bm{A}_\omega$ tends toward to $\bm{D}_\omega$,
allowing the second term in the asymptotic expansion of $\hat{\bm{\beta}}_1$ 
in equation \eqref{normal} to be neglected. 
When the Algorithm \ref{alg:paftgee} converges at the outer layer,  the penalized weighted estimator $
\hat{\bm{\beta}}_\omega$ exactly solves the estimating equation 
$\tilde{\bm{U}}_{n,\omega}(\bm{\beta},\bm{\alpha})=\bm{0}$. 
In this case, $||\bm{b}_1-\hat{\bm{\beta}}_{1,\omega}||$ is sufficiently small and the asymptotic distribution of $\hat{\bm{\beta}}_{1,\omega}$ is fully characterized by the
first term in the asymptotic expansion in \eqref{normal}, 
mirroring the asymptotic behavior of the traditional penalized BJ estimator in \citet{johnson2008penalized}. 
Even if the convergence is not achieved, 
the theoretical results indicate that the penalized weighted GEE estimator $\hat{\bm{\beta}}_\omega$ 
retains the oracle property and remains asymptotically consistent and normal at each outer-layer iteration. 
In practice, our simulation studies demonstrate that the algorithm reliably converges across all considered settings.

\section{Practical implementation details\label{sec:4}}

\subsection{Variance estimation}\label{sect:vari}

Estimating the variance of  $\bm{\hat{\beta}}_\omega$ is challenging due to the lack of a closed-form expression,
combined with the added complexity introduced by the nonrandom sampling scheme and the within-cluster correlation. 
Although bootstrap methods are commonly used when closed-form variance expressions are unavailable, 
they share similar limitations with traditional cross-validation procedures. 
In particular, resampling from the original data without accounting for the sampling design can 
produce unrepresentative bootstrap samples, leading to inaccurate variance estimates.
For this reason, we adopt the resampling procedure in \citet{jin2006least} 
for variance estimation after variable selection. 
We define the perturbed estimating equation without the penalty term by
\begin{equation}
\bm{U}_{n,\omega}^{\dagger}(\bm{\beta}, \bm{b}, \hat{\bm{\alpha}}) = \sum_{i = 1}^n(\bmX_i - \bar\bmX_\omega)^\top Z_i^\dagger \omega_i\bm{\Omega}^{-1}\{\hat{\bm{\alpha}}(\bm{b})\}\left\{\hat\bmY_{i, \omega}^\dagger(\bm{b}) - \bmX_i\bm{\beta}\right\}=\bm{0},
\notag
\end{equation}
where $Z_i^\dagger$ is a random perturbation weight with 
$\mathbb{E}(Z_i^\dagger)= \mbox{Var}(Z_i^\dagger)=1$ and is independent of the data 
$(\bm{Y}_i, \bm{\Delta}_i, \bm{X}_i), i=1,\dots, n$. 
The perturbed $\hat Y^\dagger_{ik,\omega}(\bm{b})$ and $\hat F_{\omega}^\dagger(t)$
are defined as
\begin{equation}
    \begin{split}
    &\hat Y^\dagger_{ik,\omega}(\bm{b}) = \Delta_{ik} Y_{ik}+ (1 - \Delta_{ik})\left\{ \frac{\int_{e_{ik}(\bm{b})}^\infty u \,\dif \hat{F}^\dagger_{\omega}(u)}{1 - \hat{F}^\dagger_{\omega}\{e_{ik}(\bm{b})\}} +\bmX_{ik} \bm{b}\right\}, \notag\\
    &\hat F_{\omega}^\dagger(t) = 1 - \prod_{i,k: e_{ik,\omega}(b) < t}\left[1 - \frac{Z_i^\dagger\omega_{i}\Delta_{ik}}
{\sum_{j = 1}^nZ_j^\dagger\omega_{j}\sum_{l=1}^{K}I\{e_{jl,\omega}(b)\ge e_{ik,\omega}(b)\}}\right].\notag
    \end{split}
\end{equation}
The perturbed weights $Z_i^\dagger \omega_i$ are assumed at the cluster level to avoid introducing 
unnecessary variability within clusters. 
We then apply the same estimating procedure on the perturbed estimating equation and 
denote by $\hat{\bm{\beta}}^*_\omega$ the resulting estimator of $\bm{\beta}$ under perturbation. 
The variance of $\hat{\bm\beta}_\omega$ is estimated using the empirical variance of 
a large number of perturbed realizations $\hat{\bm{\beta}}^*_\omega$.
Motivated by the asymptotic normality of the estimator, we can construct a level $(1-\alpha)$ 
Wald confidence interval for each component of $\bm{\beta}$. 
For inference on the penalized estimator, $\hat{\bm{\beta}}_\omega$, 
we refit the unpenalized model using only the predictors with nonzero coefficients 
selected by the penalization procedure. 
The resampling approach is then applied to this refitted model to estimate the variance and construct 
the corresponding $(1 - \alpha)$ Wald confidence intervals.

\subsection{Correlation parameter} \label{sect:corr}

Under the assumption of common marginal distribution among $\bm{\varepsilon}_i$, 
the working correlation matrix $\bm{\Omega}(\bm{\alpha})$ can be fully specified 
through the correlation of $\varepsilon_{ik}$ and $\varepsilon_{ik^\prime}$.
Let $\alpha_{kk^\prime}$ be
the $(k,k^\prime)$th element of $\bm{\Omega}(\bm{\alpha})$, for $k, k^\prime=1,\dots, K$.  
We define $\hat{r}_{ik}=\hat{Y}_{ik}(\hat{\bm{\beta}}) - \bm{X}\hat{\bm{\beta}}$ as the residual with respect to the estimates $\hat{\bm{\beta}}$ for the $k$th subject in the 
$i$th cluster. 
Given an initial estimator $\bm{b}$, $\alpha_{kk^\prime}$ can be estimated using a weighted general method in \cite{wang2014variable} where
$
    \hat{\alpha}_{kk^\prime} = \hat{\phi}^{-1}\left(\sum_{i=1}^n \omega_i  \hat{r}_{ik} \hat{r}_{ik^\prime}\right) \left(\sum_{i=1}^n \omega_i-p\right)^{-1}$
and $\hat{\phi}= (\sum_{i=1}^n \omega_i \sum_{k=1}^{K} \hat{r}_{ik}^2)/(\sum_{i=1}^n \omega_iK-p)$.
For a specified working covariance structure $\bm{\Omega}(\bm{\alpha})$, $\bm{\alpha}$ can be estimated using residual-based moment method proposed in \citet{liang1986longitudinal} and \citet{wang2014variable}. For example, for the exchangeable correlation structure, where $\alpha_{kk^\prime}=\alpha$ for $k \neq k^\prime$ and $\bm{\Omega}(\alpha) = (1-\alpha)\bm{I} + \alpha \bm{1}\bm{1}^\top$, $\alpha$  can be estimated with sampling weights by 
$
    \hat{\alpha}(\bm{b}) = \hat{\phi}^{-1} \left(\sum_{i=1}^n \sum_{k>k'}  \omega_i \hat{r}_{ik}\hat{r}_{ik^{'}}\right)\left(  \frac{1}{2} \sum_{i=1}^n  \omega_i K(K-1)\right)^{-1}.
$
The asymptotic properties of the estimators do not depend on the specific choice of $\bm{\alpha}$ and $\phi$, as has been demonstrated in the proofs of Theorems 1 and 2. Higher efficiency can be obtained when the choice of $\bm{\Omega}(\bm{\alpha})$ is close to the truth, 
as demonstrated in our simulation results in Section \ref{sec:5} when the censoring rate is low or moderate (e.g., $ 50\%$).

\subsection{Initial value}\label{sect:4.3}
As established in Theorems 1 and 2, 
a consistent initial value is required to guarantee the consistency of the proposed estimator.
A practical and effective choice is the Gehan’s estimator, 
obtained using the smoothed rank-based approach proposed by \citet{chiou2015rank}.
 For computationally simplicity, we also consider the weighted ordinal least-squares estimator, 
denoted by $\hat{\bm{\beta}}_{ {\mbox{WOLS}}}$, 
which is based on uncensored observations and minimizes the weighted mean squared error
$\frac{1}{n}\sum_{i=1}^n\sum_{k=1}^K\omega_i\Delta_{ik}(Y_{ik}-\bmX_{ik}\bm{\beta})^2$.
The estimator $\hat{\bm{\beta}}_{ {\mbox{WOLS}}}$ takes the form 
$\hat{\bm{\beta}}_{ {\mbox{WOLS}}}=(\tilde{\bmX}^\top\bm{W}\tilde{\bmX})^{-1}\tilde{\bmX}^\top\bm{W}\tilde{\bmY}$, 
where $\tilde{\bmX}, \tilde{\bmY}$ are the  covariate matrix and  
failure time vector for uncensored observations, and $\bm{W}$ is the corresponding diagonal matrix of sampling weights. Note that  $\hat{\bm{\beta}}_{ {\mbox{WOLS}}}$ is generally biased due to the omission of censored observations without applying adjustments such as inverse probability of censoring weights. Nevertheless, our simulation studies suggest that the proposed procedures remain satisfactory estimation and selection accuracy
even when initialized with a less ideal estimator such as $\hat{\bm{\beta}}_{ {\mbox{WOLS}}}$.

Preliminary results presented in Table 1 of the Supplementary Materials indicate that 
the proposed method produces nearly unbiased estimates, regardless of the choice of initial value. 
As shown in Table 2, the variable selection performance under SCAD regularization is also comparable 
between the two initial values. 
In addition, Figures 1 through 4 plot the pairs of proposed estimates obtained from the two initial values 
under various error distributions and censoring rates. 
The strong alignment of the estimates along the 45-degree line illustrates their close agreement across settings. 
These findings suggest that the proposed method is robust to the choice of initial value in the simulation study. 
Based on this observation, we used the weighted ordinary least-squares estimator from 
uncensored observations as the initial value in all simulations and data analyses.

\section{Simulation study\label{sec:5}}
We evaluated the performance of the proposed estimating and variable selection procedures 
through simulation studies conducted under a stratified cohort design within the multivariate AFT model framework.
Motivated by the settings in  \citet{ni2016variable}, 
we set the true regression coefficient as
$\bm{\beta}_{0}=(0.35, 0, 0, 0.6, 0, 0, -0.8, 0, 0, 0.6, 0, 0, -0.8,0,..., 0)^\top$, and the dimension of the true regression coefficient $p = 18$, with six nonzero components.  
We considered a full-cohort of size $N=3000$, with each cluster consisting of $K = 3$ individuals. 
The log-multivariate failure times, $\bm{T}_i = (T_{i1}, T_{i2}, T_{i3})^\top$,  were generated 
from the linear model $\bm{T}_i = \bm{X}_i \bm{\beta}_{0}+\bm{\varepsilon}_i$, for $i=1,\dots, N$. 
Three marginal error distributions were examined, including standard normal (SN), standard logistic (SL) and standard Gumble (SG). 
A Clayton copula was used to specify three levels of within-cluster dependency among error terms, measured by Kendall's tau ($\tau$): $0$, $0.3$, and $0.6$. 
The true covariance structure was considered to be an exchangeable structure. The covariate matrix $\bm{X}_i=(\bm{X}_{i1},\dots, \bm{X}_{ip})$, where $\bm{X}_{ij}=(X_{ij1},X_{ij2},X_{ij3})^\top$ for $j=1,\dots,p$, 
was generated independently from a multivariate standard normal, 
with the correlation between $X_{ijk_1}$ and $X_{ijk_2}$ specified as $0.5^{|k_1-k_2|}$ for $k_1,k_2=1,2,3$.
The censoring times were  independently generated for each observation from a uniform distribution over $(0, \kappa)$, where $\kappa$ was chosen to achieve different censoring rates. The censoring rates, donated as C$_r$, was measured across all observations and set to $80\%$ and $90\%$ in simulations, 
reflecting the typically high censoring in stratified case-cohort designs.
After generating the full cohort data, we mimicked a stratified cohort design by 
dividing the cohort into four strata based on the number of observed events within each cluster. 
Specifically, strata 1 through 4 included clusters with 0, 1, 2, or all 3 members experiencing an event, 
respectively. 
From each stratum, clusters were randomly sampled with stratum-specific inclusion probabilities 
$p_{l}$ for the $l$th stratum, $l=1,\dots,4$. We considered $p_1=0.1, p_2=0.2, p_3=0.3, p_4=0.6$ to prioritize the clusters with more members were observed with an event. 
The corresponding weights were 10.0, 5.1, 3.3, and 1.6 for strata 1, 2, 3, and 4, respectively. 
The resulting average stratified sample sizes were $n=400$ and $340$ for censoring rates $80\%$ and $90\%$, 
respectively.

We investigated the performance of penalized variable selection using the SCAD penalty with tunning parameter $\lambda_n^{ \mbox{CV}}$ and $\lambdaSE$. 
We compared the proposed approach with the unweighted counterpart, evaluated by setting $\omega_i = 1$ in the proposed algorithm. 
For reference, we also included oracle models fitted using the true subset of covariates. 
To evaluate the impact of the working correlation structure in GEE, 
we examined both the working independence (WI) and exchangeable (EX) structures.
The variance of the estimator was estimated from the resampling approach with the bootstrap sample sizes of $200$. 
The cut-off value $10^{-3}$ as in \citet{wang2012penalized} was used, 
and any estimated coefficient below the cut-off was considered zero.

\begin{table}[t!]
\caption{Variable selection  based on 1000 replications. The working covariance structure is exchangeable with Kendall's $\tau=0.6$. SCAD$_1$ refers to SCAD penalty under the CV rule, and SCAD$_2$ refers to SCAD under the one standard error rule. TP$/$FP are the average true$/$false positives, respectively; C is the correct rate of identifying the true model; ME is the median model error; MSE is the mean square error of the estimates. SN, SL, and SG correspond to standard normal, standard logistic, and standard Gumbel distributions, respectively.}
\centering{
\resizebox{0.99\columnwidth}{!}{%
\begin{tabular}{clrrrrrrrrrrrrrrr}
\toprule
\multicolumn{1}{c}{} & \multicolumn{1}{c}{} & \multicolumn{5}{c}{SN} & \multicolumn{5}{c}{SG} & \multicolumn{5}{c}{SL} \\
\cmidrule(l{3pt}r{3pt}){3-7} \cmidrule(l{3pt}r{3pt}){8-12} \cmidrule(l{3pt}r{3pt}){13-17}
Cr & Method & TP & FP & C(\%) & ME & MSE & TP & FP & C(\%) & ME & MSE & TP & FP & C(\%) & ME & MSE\\
\midrule
\addlinespace[0.3em]
\multicolumn{17}{l}{\textbf{Without weights}}\\
\hspace{1em} & SCAD$_1$ & 6.00 & 1.6 & 37.6 & 71.2 & 0.23 & 6.00 & 1.1 & 42.6 & 42.0 & 0.18 & 6.00 & 2.0 & 26.2 & 79.3 & 0.24\\

\hspace{1em} & SCAD$_2$ & 6.00 & 0.6 & 59.4 & 81.0 & 0.25 & 6.00 & 0.2 & 86.1 & 47.5 & 0.19 & 6.00 & 0.8 & 49.1 & 91.2 & 0.26\\

\multirow[t]{-3}{*}{\raggedleft\arraybackslash 80\%} & Oracle &  6.00 & 0.0 & 100.0 & 62.0 & 0.21 & 6.00 & 0.0 & 100.0 & 38.5 & 0.17 & 6.00 & 0.0 & 100.0 & 71.1 & 0.23\\
\addlinespace[0.3em]
\hspace{1em} & SCAD$_1$ & 5.97 & 1.2 & 43.2 & 101.8 & 0.31 & 5.97 & 0.2 & 77.7 & 57.3 & 0.24 & 5.96 & 1.6 & 36.1 & 127.7 & 0.34\\

\hspace{1em} & SCAD$_2$ & 5.95 & 0.5 & 66.3 & 116.0 & 0.33 & 5.93 & 0.2 & 77.6 & 66.8 & 0.26 & 5.94 & 0.8 & 55.3 & 139.2 & 0.36\\

\multirow[t]{-3}{*}{\raggedleft\arraybackslash 90\%} & Oracle & 6.00 & 0.0 & 100.0 & 78.9 & 0.27 & 6.00 & 0.0 & 100.0 & 48.9 & 0.22 & 6.00 & 0.0 & 100.0 & 94.0 & 0.30\\
\cmidrule{1-17}
\addlinespace[0.3em]
\multicolumn{17}{l}{\textbf{With weights}}\\
\hspace{1em} & SCAD$_1$ & \multirow[t]{-1}{*}{\raggedleft\arraybackslash 6.00} & 0.9 & 48.0 & 27.8 & 0.16 & 6.00 & 0.4 & 69.8 & 18.5 & 0.13 & 6.00 & 0.9 & 46.4 & 28.7 & 0.16\\

\hspace{1em} & SCAD$_2$ & 5.99 & 0.2 & 85.5 & 35.6 & 0.18 & 6.00 & 0.2 & 80.4 & 21.4 & 0.14 & 6.00 & 0.1 & 90.2 & 39.1 & 0.19\\

\multirow[t]{-3}{*}{\raggedleft\arraybackslash 80\%} & Oracle & 6.00 & 0.0 & 100.0 & 24.4 & 0.14 & 6.00 & 0.0 & 100.0 & 18.5 & 0.13 & 6.00 & 0.0 & 100.0 & 26.5 & 0.15\\
\addlinespace[0.3em]
\hspace{1em} & SCAD$_1$ & 5.89 & 0.2 & 75.0 & 56.7 & 0.25 & 5.90 & 0.0 & 88.6 & 39.6 & 0.22 & 5.88 & 0.3 & 67.5 & 73.1 & 0.28\\

\hspace{1em} & SCAD$_2$ & 5.80 & 0.1 & 68.8 & 68.9 & 0.28 & 5.84 & 0.0 & 82.3 & 42.0 & 0.23 & 5.78 & 0.2 & 61.4 & 95.1 & 0.31\\

\multirow[t]{-3}{*}{\raggedleft\arraybackslash 90\%} & Oracle & 6.00 & 0.0 & 100.0 & 47.5 & 0.22 & 6.00 & 0.0 & 100.0 & 35.1 & 0.19 & 6.00 & 0.0 & 100.0 & 54.4 & 0.24\\
\bottomrule
\end{tabular}
}}
\label{multi:pen1}
\end{table}

\begin{table}[t!]
\caption{Estimation performance for $\beta_{1}=0.35$ using the 1000 replications that correctly identified $\beta_{01}$ in the model. The working covariance structure is exchangeable with Kendall's $\tau=0.6$.  SCAD$_1$ refers to SCAD penalty under the CV rule, and SCAD$_2$ refers to SCAD penalty under the one standard error rule.  $N_c$ is the number of replications where $\hat{\beta}_{1} \neq 0$; BR is the bias ratio as a percentage; SE$_a/$SE$_e$  $(\times 10^{-2})$ are averaged$/$empirical standard error of the estimator, respectively; CP is 95$\%$ empirical confidence interval coverage probability. }
\centering{
\resizebox{0.99\columnwidth}{!}{%
\begin{tabular}{clrrrrrrrrrrrrrrr}
\toprule
\multicolumn{1}{c}{} & \multicolumn{1}{c}{} & \multicolumn{5}{c}{SN} & \multicolumn{5}{c}{SG} & \multicolumn{5}{c}{SL} \\
\cmidrule(l{3pt}r{3pt}){3-7} \cmidrule(l{3pt}r{3pt}){8-12} \cmidrule(l{3pt}r{3pt}){13-17}
& & $N_c$ & BR & SE$_a$ & SE$_e$ & CP & $N_c$ & BR & SE$_a$ & SE$_e$ & CP & $N_c$ & BR & SE$_a$ & SE$_e$ & CP\\
Cr & Method&&(\%)&&&(\%)&&(\%)&&&(\%)&&(\%)&&&(\%)\\
\midrule
\addlinespace[0.3em]
\multicolumn{17}{l}{\textbf{Without weights}}\\
\hspace{1em} & SCAD$_1$ & 1000 & 17.9 & 4.25 & 5.66 & 78 & 1000 & 12.1 & 3.64 & 4.39 & 82 & 1000 & 17.8 & 4.46 & 5.73 & 80\\

\hspace{1em} & SCAD$_2$ & 1000 & 25.5 & 4.26 & 7.02 & 78 & 1000 & 19.2 & 3.66 & 5.85 & 82 & 1000 & 25.9 & 4.46 & 6.78 & 82\\

\multirow[t]{-3}{*}{\raggedleft\arraybackslash 80\%} & Oracle & 1000 & 12.1 & 4.25 & 4.32 & 82 & 1000 & 9.3 & 3.70 & 3.86 & 84 & 1000 & 13.3 & 4.46 & 4.40 & 81\\
\addlinespace[0.3em]
\hspace{1em} & SCAD$_1$ & 971 & 23.2 & 6.61 & 10.89 & 86 & 970 & 14.4 & 5.59 & 8.67 & 86 & 961 & 26.0 & 7.19 & 11.65 & 85\\

\hspace{1em} & SCAD$_2$ & 954 & 33.6 & 6.61 & 12.71 & 86 & 935 & 20.5 & 5.62 & 11.24 & 88 & 943 & 35.5 & 7.21 & 12.86 & 87\\

\multirow[t]{-3}{*}{\raggedleft\arraybackslash 90\%} & Oracle & 1000 & 13.6 & 6.59 & 6.85 & 86 & 1000 & 11.1 & 5.62 & 5.97 & 87 & 1000 & 13.0 & 7.24 & 7.52 & 88\\
\cmidrule{1-17}
\addlinespace[0.3em]
\multicolumn{17}{l}{\textbf{With weights}}\\
\hspace{1em} & SCAD$_1$ & \multirow[t]{-1}{*}{\raggedleft\arraybackslash 1000} & 5.1 & 5.39 & 7.24 & 92 & 1000 & 3.3 & 4.65 & 5.56 & 94 & 1000 & 3.7 & 5.56 & 7.02 & 93\\

\hspace{1em} & SCAD$_2$ & 993 & 17.1 & 5.40 & 10.89 & 92 & 997 & 8.8 & 4.66 & 7.57 & 93 & 995 & 19.4 & 5.59 & 11.41 & 93\\

\multirow[t]{-3}{*}{\raggedleft\arraybackslash 80\%} & Oracle & 1000 & 0.5 & 5.41 & 5.54 & 94 & 1000 & 0.7 & 4.71 & 4.98 & 93 & 1000 & 1.5 & 5.56 & 5.49 & 95\\
\addlinespace[0.3em]
\hspace{1em} & SCAD$_1$ & 896 & 5.5 & 7.78 & 11.28 & 93 & 901 & 4.3 & 6.49 & 8.78 & 93 & 884 & 4.8 & 8.40 & 12.20 & 93\\

\hspace{1em} & SCAD$_2$ & 805 & 6.7 & 7.84 & 12.33 & 94 & 836 & 3.1 & 6.50 & 8.75 & 94 & 783 & 7.9 & 8.43 & 13.89 & 93\\

\multirow[t]{-3}{*}{\raggedleft\arraybackslash 90\%} & Oracle & 1000 & 2.5 & 7.72 & 8.19 & 94 & 1000 & 4.3 & 6.50 & 7.10 & 92 & 1000 & 0.0 & 8.39 & 8.58 & 94\\
\bottomrule
\end{tabular}
}}
\label{multi:pen2}
\end{table}

The performance of variable selection was evaluated using three summary measures: 
true positives (TP), calculated as the proportion of correctly identified nonzero coefficients; 
false positives (FP), calculated as the proportion of incorrectly selected zero coefficients; 
and correct rate (C), calculated as the proportion of simulations in which the true model was exactly recovered. 
These results are presented in Table \ref{multi:pen1}.
Additionally, we reported the median model error (ME), given by the median of 
$\sum_{i=1}^{n}\{\sum_{j=1}^{18} (\hat{\beta}_j -{\beta}_{0j}) X_{ji}\}^2$, 
where $\hat{\beta}_j$ is the penalized (weighted or unweighted) estimator of $\beta_{0j}$, 
and the MSE of $\hat{\bm{\beta}}$, as the average of $\sum_{j=1}^{18}(\hat{\beta}_j - \beta_{0j})^2$. 
To examine the estimation accuracy after variable selection, 
we focused on the results for $\beta_{01}=0.35$, 
which has the smallest effect size among the nonzero coefficients and 
is therefore the most likely to be penalized toward zero.
Table~\ref{multi:pen2} summarizes the bias ratio (BR), empirical standard error (SE$_e$), averaged standard error (SE$_a$), mean square error (MSE), and $95\%$ empirical confidence interval coverage probability (CP)  for $\hat{\beta}_{1}$, using the replications that correctly identified $\beta_{01}$ as
non-zero. 
We define the BR as the average percentage of $(\hat{\beta_1}-\beta_{01})/\beta_{01}$,
SE$_e$ as the standard error of the estimates,
SE$_a$ as the average of the estimated standard deviation from the multiplier resampling approach,
and CP as the proportion of the constructed $95\%$ Wald confidence intervals containing the true $\beta_{01}$. The value of $N_c$ is the number of replications where $\hat{\beta}_{1} \neq 0$.

Table \ref{multi:pen1} and \ref{multi:pen2} focused on a high censoring rate with the EX structure and compared the effectiveness of the weighted approaches in variable selection with their unweighted counterparts. Firstly, both the weighted and unweighted approaches achieved fairly high TP values, 
indicating that both approaches correctly identified all nonzero coefficients in the majority of replications. 
This suggests that even under substantial censoring, 
both methods maintain strong sensitivity in detecting relevant covariates.
At the same time, the weighted approaches consistently produced fewer false positives and 
higher correct model selection rates than their unweighted counterparts across all 
three error distributions and both high censoring levels. 
This highlights the advantage of incorporating sampling weights in improving variable selection accuracy 
by reducing false discoveries and enhancing recovery of the true model.
Secondly, Table \ref{multi:pen1}  demonstrates that the weighted approach with tuning parameter 
$\lambdaSE$ substantially reduces both ME and MSE compared to the unweighted approach, 
while maintaining satisfactory variable selection performance. 
In addition, Table \ref{multi:pen2} shows that the unweighted approach resulted in large BR 
and poor coverage probabilities, 
whereas the weighted approach achieved coverage probabilities close to the nominal 95\% level.
The results for ME and MSE in Table~\ref{multi:pen1}, 
together with BR and CP in Table~\ref{multi:pen2}, 
suggest that incorporating sampling weights into the penalized GEE framework improves estimation accuracy 
after variable selection, even under high levels of censoring.
Finally, Table \ref{multi:pen1} shows that the penalized GEE under SCAD$_2$ (with $\lambdaSE$) 
yields more favorable FP and C compared to SCAD$_1$ (with $\lambdaCV$), although it exhibited larger ME and MSE. 
This observation highlights a trade-off between estimation precision and 
variable selection performance depending on the choice of tuning parameter.

We further compared the performance when fitting the WI and EX models under moderate and high censoring rates in Table \ref{multi:pen3} and \ref{multi:pen4}. We observed substantial improvements in the ME, MSE, BR, and CP by incorporating weights under both WI and EX structures and each level of within-cluster dependency. 
The SE$_a$ of estimates is close to the SE$_e$ for both weighted and unweighted approaches for moderate censoring (i.e. $50\%$), indicating that the resampling procedure yields a valid variance estimation.
Specifically, from Table \ref{multi:pen3}, in the setting of $50\%$ censoring and non-zero Kendall's tau, the EX model reduces the ME and MSE without compromising variable selection performance compared to the WI model, and the reduction becomes more evident as the level of within-cluster dependence increases. 
Additionally, from Table \ref{multi:pen4}, the estimated variances of the estimates under the EX structure are generally smaller than those under the WI structure when the censoring rate is $50\%$, and Kendall's tau is non-zero.
This is anticipated as the true covariance structure in this simulation setting is
exchangeable. However, we did not observe a similar trend of reduced variance with the EX structure when censoring
was at $90\%$. This may be attributed to the high level of censoring diluting the within-cluster dependence, leading to
a weaker signal for the EX model.
In contrast, when the censoring rate is as high as 90\%, 
using the EX structure in penalized GEE showed no improvement in ME, MSE, and estimated variance compared to using the WI structure.
Lastly, it is worth noting that improvements in estimation and variable selection may not be achieved simultaneously.  
For example, when the censoring rate is $90\%$, holding the level of $\tau$, the EX structure improved the variable selection in terms of FP and C values, while increasing the model errors compared with WI structure.  In situations where the variable selection performance takes priority over estimation accuracy,
the WI structure could be considered in variable selection when the censoring rate is high (i.e. 90\%).  
Nevertheless, the difference in variable selection and estimation performance between the WI and EX structures is still relatively small, echoing that the proposed penalized GEE is robust to the misspecification of the working covariance structure.

\begin{table}[t!]
\caption{Variable selection based on 1000 replications. The marginal distribution of error terms is standard normal. WI/EX are working independent/exchangeable structure; SCAD$_1$ refers to SCAD penalty under the CV rule, and SCAD$_2$ refers to SCAD under the one standard error rule. TP$/$FP are the average true$/$false positives, respectively; C is the correct rate of identifying the true model; ME is the median model error; MSE is the mean square error of the estimates.}
\begin{center}
\resizebox{0.99\columnwidth}{!}{%
\begin{tabular}{crlrrrrrrrrrrrrrrr}
\toprule
\multicolumn{1}{c}{} & \multicolumn{1}{c}{} & \multicolumn{1}{c}{} & \multicolumn{5}{c}{Unweighted-WI} & \multicolumn{5}{c}{Weighted-WI} & \multicolumn{5}{c}{Weighted-EX} \\
\cmidrule(l{3pt}r{3pt}){4-8} \cmidrule(l{3pt}r{3pt}){9-13} \cmidrule(l{3pt}r{3pt}){14-18}
Cr & $\tau$ & Method & TP & FP & C(\%) & ME & MSE & TP & FP & C(\%) & ME & MSE & TP & FP & C(\%) & ME & MSE\\
\midrule
 &  & SCAD$_1$ & 6.00 & 2.2 & 33.6 & 46.9 & 0.16 & 6.00 & 1.4 & 47.7 & 14.9 & 0.09 & 6.00 & 1.4 & 50.0 & 15.0 & 0.09\\

 &  & SCAD$_2$ & 6.00 & 0.0 & 95.6 & 73.1 & 0.20 & 6.00 & 0.0 & 98.3 & 27.1 & 0.12 & 6.00 & 0.0 & 98.4 & 27.8 & 0.13\\

 & \multirow[t]{-3}{*}{\raggedleft\arraybackslash 0.0} & Oracle & 6.00 & 0.0 & 100.0 & 44.4 & 0.15 & 6.00 & 0.0 & 100.0 & 12.3 & 0.08 & 6.00 & 0.0 & 100.0 & 12.1 & 0.08\\
\addlinespace[0.3em]
 &  & SCAD$_1$ & 6.00 & 2.0 & 22.4 & 59.6 & 0.18 & 6.00 & 1.8 & 40.8 & 17.3 & 0.10 & 6.00 & 1.4 & 44.8 & 15.6 & 0.09\\

 &  & SCAD$_2$ & 6.00 & 0.2 & 82.6 & 94.9 & 0.22 & 6.00 & 0.1 & 95.6 & 28.0 & 0.13 & 6.00 & 0.0 & 98.1 & 28.0 & 0.13\\

 & \multirow[t]{-3}{*}{\raggedleft\arraybackslash 0.3} & Oracle & 6.00 & 0.0 & 100.0 & 54.4 & 0.17 & 6.00 & 0.0 & 100.0 & 13.8 & 0.09 & 6.00 & 0.0 & 100.0 & 13.0 & 0.09\\
\addlinespace[0.3em]
 &  & SCAD$_1$ & 6.00 & 2.4 & 13.7 & 63.6 & 0.18 & 6.00 & 2.1 & 33.8 & 18.3 & 0.10 & 6.00 & 2.1 & 32.9 & 15.2 & 0.09\\

 &  & SCAD$_2$ & 6.00 & 0.3 & 73.3 & 94.9 & 0.22 & 6.00 & 0.1 & 92.9 & 28.3 & 0.13 & 6.00 & 0.1 & 93.9 & 22.7 & 0.11\\

\multirow[t]{-9}{*}{\raggedleft\arraybackslash 50\%} & \multirow[t]{-3}{*}{\raggedleft\arraybackslash 0.6} & Oracle & 6.00 & 0.0 & 100.0 & 55.5 & 0.17 & 6.00 & 0.0 & 100.0 & 14.7 & 0.09 & 6.00 & 0.0 & 100.0 & 13.3 & 0.09\\
\cmidrule{1-18}
 &  & SCAD$_1$ & 5.97 & 1.0 & 37.7 & 69.7 & 0.27 & 5.89 & 0.2 & 74.0 & 52.7 & 0.24 & 5.90 & 0.2 & 70.4 & 52.2 & 0.24\\

 &  & SCAD$_2$ & 5.92 & 0.6 & 57.6 & 89.4 & 0.29 & 5.80 & 0.2 & 67.9 & 66.8 & 0.27 & 5.82 & 0.2 & 66.4 & 61.9 & 0.27\\

 & \multirow[t]{-3}{*}{\raggedleft\arraybackslash 0.0} & Oracle & 6.00 & 0.0 & 100.0 & 55.5 & 0.23 & 6.00 & 0.0 & 100.0 & 42.1 & 0.21 & 6.00 & 0.0 & 100.0 & 43.4 & 0.21\\
\addlinespace[0.3em]
 &  & SCAD$_1$ & 5.96 & 1.4 & 33.0 & 80.8 & 0.28 & 5.92 & 0.3 & 69.0 & 50.8 & 0.24 & 5.89 & 0.1 & 77.1 & 54.0 & 0.25\\

 &  & SCAD$_2$ & 5.94 & 0.8 & 55.5 & 96.1 & 0.30 & 5.83 & 0.2 & 65.3 & 63.8 & 0.27 & 5.79 & 0.1 & 69.1 & 68.4 & 0.27\\

 & \multirow[t]{-3}{*}{\raggedleft\arraybackslash 0.3} & Oracle & 6.00 & 0.0 & 100.0 & 56.7 & 0.24 & 6.00 & 0.0 & 100.0 & 42.6 & 0.21 & 6.00 & 0.0 & 100.0 & 44.0 & 0.21\\
\addlinespace[0.3em]
 &  & SCAD$_1$ & 5.96 & 1.5 & 34.5 & 81.2 & 0.28 & 5.90 & 0.3 & 69.1 & 48.6 & 0.24 & 5.89 & 0.2 & 75.0 & 56.7 & 0.25\\

 &  & SCAD$_2$ & 5.95 & 0.8 & 57.1 & 89.5 & 0.29 & 5.82 & 0.2 & 65.9 & 62.0 & 0.26 & 5.80 & 0.1 & 68.8 & 68.9 & 0.28\\

\multirow[t]{-9}{*}{\raggedleft\arraybackslash 90\%} & \multirow[t]{-3}{*}{\raggedleft\arraybackslash 0.6} & Oracle & 6.00 & 0.0 & 100.0 & 54.6 & 0.23 & 6.00 & 0.0 & 100.0 & 44.1 & 0.21 & 6.00 & 0.0 & 100.0 & 47.5 & 0.22\\
\bottomrule
\end{tabular}
}
\end{center}
\label{multi:pen3}
\end{table}

\begin{table}[t!]
\caption{Estimation performance for $\beta_1=0.35$ using 1000 replications that correctly identified $\beta_{01}$ in the model. The marginal distribution of error terms is standard normal.  SCAD$_1$ refers to SCAD penalty under the CV rule, and SCAD$_2$ refers to SCAD penalty under the one standard error rule.  $N_c$ is the number of replications where $\hat{\beta}_{1} \neq 0$; BR is the bias ratio as a percentage; SE$_a/$SE$_e$  $(\times 10^{-2})$ are averaged$/$empirical standard error of the estimator, respectively; CP is 95$\%$ empirical confidence interval coverage probability.}
\centering{
\resizebox{0.99\columnwidth}{!}{%
\begin{tabular}{crlrrrrrrrrrrrrrrr}
\toprule
\multicolumn{1}{c}{} & \multicolumn{1}{c}{} & \multicolumn{1}{c}{} & \multicolumn{5}{c}{Unweighted-WI} & \multicolumn{5}{c}{Weighted-WI} & \multicolumn{5}{c}{Weighted-EX} \\
\cmidrule(l{3pt}r{3pt}){4-8} \cmidrule(l{3pt}r{3pt}){9-13} \cmidrule(l{3pt}r{3pt}){14-18}
&&  & $N_c$ & BR & SE$_a$ & SE$_e$ & CP & $N_c$ & BR & SE$_a$ & SE$_e$ & CP & $N_c$ & BR & SE$_a$ & SE$_e$ & CP\\
Cr & $\tau$ & Method&&(\%)&&&(\%)&&(\%)&&&(\%)&&(\%)&&&(\%)\\
\midrule
 &  & SCAD$_1$ & 1000 & 10.5 & 2.84 & 3.35 & 79 & 1000 & 2.7 & 3.41 & 4.13 & 92 & 1000 & 3.4 & 3.42 & 3.88 & 94\\

 &  & SCAD$_2$ & 1000 & 29.3 & 2.85 & 4.03 & 78 & 1000 & 24.0 & 3.42 & 4.89 & 93 & 1000 & 24.5 & 3.44 & 4.67 & 94\\

 & \multirow[t]{-3}{*}{\raggedleft\arraybackslash 0.0} & Oracle & 1000 & 8.4 & 2.85 & 2.98 & 81 & 1000 & -0.2 & 3.42 & 3.55 & 93 & 1000 & 0.2 & 3.41 & 3.47 & 95\\
\addlinespace[0.3em]
 &  & SCAD$_1$ & 1000 & 12.6 & 3.27 & 4.13 & 80 & 1000 & 3.7 & 3.58 & 4.40 & 92 & 1000 & 3.3 & 3.39 & 4.27 & 93\\

 &  & SCAD$_2$ & 1000 & 29.8 & 3.28 & 4.52 & 80 & 1000 & 24.1 & 3.59 & 5.03 & 93 & 1000 & 24.9 & 3.41 & 4.99 & 93\\

 & \multirow[t]{-3}{*}{\raggedleft\arraybackslash 0.3} & Oracle & 1000 & 9.5 & 3.29 & 3.34 & 82 & 1000 & 0.3 & 3.62 & 3.67 & 95 & 1000 & 0.7 & 3.42 & 3.49 & 94\\
\addlinespace[0.3em]
 &  & SCAD$_1$ & 1000 & 12.5 & 3.36 & 4.14 & 82 & 1000 & 2.6 & 3.72 & 4.37 & 93 & 1000 & 3.8 & 3.30 & 3.90 & 93\\

 &  & SCAD$_2$ & 1000 & 28.5 & 3.38 & 4.47 & 81 & 1000 & 22.6 & 3.76 & 5.12 & 93 & 1000 & 19.9 & 3.32 & 4.44 & 93\\

\multirow[t]{-9}{*}{\raggedleft\arraybackslash 50\%} & \multirow[t]{-3}{*}{\raggedleft\arraybackslash 0.6} & Oracle & 1000 & 9.4 & 3.40 & 3.53 & 83 & 1000 & 0.2 & 3.77 & 3.86 & 94 & 1000 & 0.1 & 3.32 & 3.33 & 94\\
\cmidrule{1-18}
 &  & SCAD$_1$ & 968 & 15.4 & 6.73 & 10.81 & 89 & 891 & 4.6 & 7.58 & 10.86 & 95 & 897 & 4.9 & 7.61 & 10.68 & 94\\

 &  & SCAD$_2$ & 921 & 25.0 & 6.76 & 13.30 & 92 & 807 & 8.0 & 7.55 & 12.64 & 96 & 824 & 8.0 & 7.63 & 12.44 & 94\\

 & \multirow[t]{-3}{*}{\raggedleft\arraybackslash 0.0} & Oracle & 1000 & 10.1 & 6.81 & 7.41 & 88 & 1000 & 2.3 & 7.56 & 8.16 & 92 & 1000 & 1.2 & 7.66 & 8.29 & 93\\
\addlinespace[0.3em]
 &  & SCAD$_1$ & 964 & 15.5 & 7.18 & 11.45 & 92 & 921 & 4.8 & 7.58 & 10.33 & 95 & 892 & 4.7 & 7.68 & 11.15 & 94\\

 &  & SCAD$_2$ & 946 & 26.4 & 7.24 & 13.99 & 94 & 832 & 8.4 & 7.58 & 12.46 & 95 & 795 & 6.8 & 7.64 & 12.02 & 95\\

 & \multirow[t]{-3}{*}{\raggedleft\arraybackslash 0.3} & Oracle & 1000 & 8.7 & 7.37 & 8.16 & 90 & 1000 & 2.3 & 7.68 & 8.20 & 93 & 1000 & 1.0 & 7.67 & 8.18 & 94\\
\addlinespace[0.3em]
 &  & SCAD$_1$ & 965 & 15.0 & 7.33 & 11.92 & 90 & 904 & 3.5 & 7.65 & 10.64 & 93 & 896 & 5.5 & 7.78 & 11.28 & 93\\

 &  & SCAD$_2$ & 949 & 23.3 & 7.40 & 14.03 & 91 & 820 & 5.9 & 7.64 & 12.27 & 94 & 805 & 6.7 & 7.84 & 12.33 & 94\\

\multirow[t]{-9}{*}{\raggedleft\arraybackslash 90\%} & \multirow[t]{-3}{*}{\raggedleft\arraybackslash 0.6} & Oracle & 1000 & 6.7 & 7.45 & 8.20 & 89 & 1000 & 1.2 & 7.70 & 8.27 & 94 & 1000 & 2.5 & 7.72 & 8.19 & 94\\
\bottomrule
\end{tabular}
}}
\label{multi:pen4}
\end{table}

\section{Matched Cohort Dental Study\label{sec:6}}
Deep caries or restorations can result in pulpal involvement, 
often requiring RC treatment or tooth extraction. 
While RC treatment may extend the life of a tooth, it does not necessarily prevent future tooth loss. 
We applied the proposed methods to a retrospective cohort dental study \citep{caplan2005root} to evaluate the impact of pulpal involvement and subsequent RC treatment on tooth survival, 
as well as to assess the effectiveness of RC treatment across different tooth types. 
The data were obtained from current or retired employees and 
their dependents of companies with dental insurance through the Kaiser Permanente Dental Care Program, 
a dental health maintenance organization located in Portland, Oregon, during the period from 1987 to 1994.
The full cohort consisted of a total of 1,795 patients who had at least one RC-filled tooth.
Case and control patients were defined as those who lost the RC-filled tooth during the study period and those who did not, respectively. 
In this study, we had 272 cases and 1,523 controls. 
RC treatment was used as a proxy for pulpal involvement and served as the time origin for measuring tooth survival.
Given the slow disease progression, stratified case-cohort sampling was used to sample from two strata consisting of all cases and all controls, respectively.
Patients were randomly sampled within cases and controls using different inclusion probabilities of $85.29\%$ and $11.42\%$, respectively. 
Each patient contributed with an RC-filled tooth and a contralateral non-RC-filled tooth, forming a cluster of size two. 
If the contralateral non-RC-filled tooth was missing or was also an RC-filled tooth, a tooth of the same type (i.e., anterior, premolar, or molar) adjacent to the contralateral tooth was selected. 
A total of 202 eligible patients (111 case patients and 91 control patients) were sampled after considering the study eligibility criteria in \citep{caplan2005root}. 
Complete covariate information was only available for the observations in the case-cohort, and the full cohort data are not available in this study.

We included two main risk factors, the RC treatment status (ROOT) and the tooth type (MOLAR), 
along with an interaction term (MOLAR:ROOT) to evaluate the effect of RCT on different tooth types. 
We also considered additional risk factors, including proximal contacts (PC), 
the number of pockets larger than 5 mm pocket depths (POCKET), 
the number of decayed or filled coronal surfaces (CORDF), 
the number of decayed or filled root surfaces (RTDF),  and filled surfaces (SFILL).
PC measures the contact between the distal surface of a tooth and the mesial surface of an adjacent tooth. Four mutually exclusive categories were created for PC: 
nonbridge abutment with zero PC (PC0), nonbridge abutment with one PC (PC1), 
nonbridge abutment with two PCs (PC2), 
and bridge abutment (PCABUT).
The variable CORDF is the number of decayed or filled
(DF) coronal surfaces among the occlusal, mesial, and distal coronal surfaces, with possible integer values between 0 to 3. 
We include CORDF using three dummy variables (CORDF1, CORDF2, CORDF3), corresponding to the three values of CORDF: 1, 2, and 3, respectively. 
The variable RTDF is the number of decayed or filled coronal surfaces from among the mesial and distal root surfaces, ranging from 0 to 2. Two dummy variables (RTDF1, RTDF2) were used for RTDF, corresponding to the value $1,2$ of RTDF, respectively. 
Each level of CORDF or RTDF is treated independently in the analysis, rather than applying a group selection approach to these variables.

\begin{table}[t!]
\caption{The risk factors identified by proposed variable selection procedure for matched cohort dental study. Est is the estimated coefficient; SE$_a$ is the average of the standard error of the estimates; WI$/$EX are the working independence$/$exchangeable structures, respectively. The results are based on the tuning parameter $\lambdaSE$. The covariates in unpenalized models marked with $\ast$ were considered as significant. 
}
 \centering
 \resizebox{0.88\columnwidth}{!}{%
  \renewcommand{\arraystretch}{0.7} 
\begin{tabular}{lrrrrrr}
\toprule
\multicolumn{1}{c}{} & \multicolumn{2}{c}{Without penalties} & \multicolumn{2}{c}{Lasso} & \multicolumn{2}{c}{SCAD}\\
\cmidrule(l{3pt}r{3pt}){2-3} \cmidrule(l{3pt}r{3pt}){4-7}
\multicolumn{1}{c}{} & \multicolumn{1}{c}{WI} & \multicolumn{1}{c}{EX} & \multicolumn{1}{c}{WI} & \multicolumn{1}{c}{EX}& \multicolumn{1}{c}{WI} & \multicolumn{1}{c}{EX} \\
\cmidrule(l{3pt}r{3pt}){2-2} \cmidrule(l{3pt}r{3pt}){3-3}
\cmidrule(l{3pt}r{3pt}){4-4}
\cmidrule(l{3pt}r{3pt}){5-5}
\cmidrule(l{3pt}r{3pt}){6-6}
\cmidrule(l{3pt}r{3pt}){7-7}
Variables & EST (SE) & EST (SE) &EST (SE) &EST (SE)&EST (SE)&EST (SE)\\
\midrule
MOLAR & 2.14 (1.17) & 2.12$^\ast$  (1.02) &2.25 (0.96)  & 2.05 (0.97) & 2.25 (1.08)  & 2.07 (1.00)\\
ROOT & -0.41 (0.71) & -0.31 (0.60)& -0.41 (0.64) & -0.39 (0.60) & -0.41 (0.71)   & -0.36 (0.65)\\
PC1 & 1.37 (0.89) & 1.14 (0.83)& 1.68 (0.85) & 1.39 (0.92) & 1.68 (0.99)   & 1.06 (0.63)\\
PC2 & 2.45$^\ast$ (0.86) & 1.96$^\ast$ (0.83) & 2.81 (0.80)  & 2.22 (0.86) & 2.81 (0.92)  & 1.81 (0.57)\\
PCABUT & 0.90 (1.11) & 0.62 (1.06)& 1.22 (1.05)   & 0.79 (1.05) & 1.22 (1.05)   & -\\
POCKET & -0.27 (0.19) & -0.23 (0.17)& -     & - & - & -\\
SFILL & -0.08 (0.34) & -0.02 (0.31)& -     & - & - & -\\
CORDF1 & 1.29 (1.15) & 0.48 (1.15) & 1.05 (0.94)  & 0.77 (0.87) & 1.06 (0.95)    & -\\
CORDF2 & 0.83 (0.86) & 0.10 (1.09) & -  & - & - & -\\
CORDF3 & 0.37 (0.99) & -0.35 (1.26) & -  & - & - & -\\
RTDF1 & -0.25 (0.43) & -0.29 (0.36) & - & - & - & -\\
RTDF2 & -0.39 (0.76) & -0.39 (0.66) & - & - & - & -\\
MOLAR:ROOT & -2.09$^\ast$ (1.05) & -2.08$^\ast$ (0.95) & -2.19 (0.97) & -2.05 (0.92) & -2.19 (1.09)  & -2.12 (1.01)\\
\bottomrule
\end{tabular}
}
\label{dental:nop}
\end{table}

We employed two variable selection methods, Lasso and SCAD, using $\lambdaSE$ as the tuning parameter.
For all models, the main risk factors (ROOT and MOLAR) were excluded from penalization in all models. Standard errors for the penalized estimates were obtained through the resampling procedure following variable selection. As a benchmark, we also considered weighted WI and EX models without penalization, with the standard errors similarly estimated via resampling. In the unpenalized models, the significant covariates were identified using $95\%$ Wald confidence interval  constructed based on the estimated coefficients (EST) and their corresponding standard errors (SE).

Table \ref{dental:nop} presents the risk factors identified by the unpenalized and penalized weighted GEE for AFT model. Under the unpenalized weighted WI and EX models, both PC2 and MOLAR:ROOT were found to be significantly different from zero, with the EX model indicating that the MOLAR effect was also statistically significant. 
The estimates of the risk factors identified by the penalized model are close to those obtained from the unpenalized model. 
The positive coefficient estimates of MOLAR imply that molars tend to last longer than non-molar teeth, while the negative small magnitude of coefficient estimates of ROOT indicate that the effect of RC treatment among non-molars was not significant.
All regularization methods identified PC1, PC2, and MOLAR:ROOT as significant risk factors for tooth survival. 
The positive coefficients of PC1 and PC2 suggest that teeth with one or two proximal contacts tend to have longer lifespans compared to teeth with no proximal contact. Similarly, the positive coefficients of CORDF1 indicate that teeth with one DF coronal surface tend to survive longer than those with no decayed or filled surfaces. Further, the ROOT effect on MOLAR can be determined by summing the coefficients of ROOT and MOLAR:ROOT. 
For example, the RC treatment effect on MOLAR is calculated to be $-2.60$ under WI model, suggesting that the life span of RC-filled molars is approximated 
by a factor of $\exp(2.6)$ or twelve times shorter than non-RC filled molars, such that RC filled molars have a much higher risk of extraction than non-RC filled molars, which is consistent with the findings in \citet{chiou2015semiparametric} and \citet{kang2009marginal-2}.

\section{Discussion\label{sec:7}}

We proposed a penalized weighted GEE approach for variable selection in 
semiparametric multivariate AFT models under stratified sampling designs. 
The method addresses key challenges including informative sampling, censoring, and within-cluster dependence. 
The proposed method can be readily extended to accommodate settings in which the error terms within a cluster have different marginal distributions. 
In such cases, we recommend estimating each distinct marginal distribution using the marginal weighted KM estimator, 
similar to that discussed in \citet{chiou2014marginal}. 
Using similar arguments to those used in proving the consistency of 
the pooled weighted KM estimator in Lemma 1 in the Appendix, 
the marginal estimators can also be shown to be consistent, thereby preserving the consistency of the parameter estimators. 
However, identifying dependent error terms that share the same marginal distribution remains an open question and requires further investigation.

The weighting feature of the proposed weighted least-squares approach 
is flexible and can be adapted to various study designs, 
including the nested case-control design \citep{kang2017fitting} for example. 
It can also be extended to handle length-biased sampling by incorporating the 
doubly weighted approach used in rank-based AFT models \citep{chiou2017rank}. 
Instead of using cross-validation for selecting the tuning parameter, 
alternative criteria such as the Bayesian information criterion, quasi-likelihood information criterion, and covariance inflation criterion could be considered. 
Additionally, boosting methods and gradient descent-based techniques for variable selection via estimating equations \citep{wolfson2011eeboost, chen2023noising} may be particularly useful when the number of covariates exceeds the number of clusters. 
These extensions open up promising directions for future work.



\section*{Appendix}

\begin{lemma}\label{lem:wKM}
If conditions (C1) -- (C6) hold, assume $\bm{\beta}$ and $\bm{\beta}^\prime$ are
any two estimators that in some neighborhood of $\bm{\beta}_0$ satisfying that both $\|\bm{\beta}\|<B$ and $\|\bm{\beta}'\|<B$, the weighted pooled Kaplan-Meier estimator $\hat{F}_{n,\omega}^{\bm{\beta}}(u)$ defined in equation as \eqref{km} satisfies
\begin{equation}
    \sup_{\|\bm{\beta}\|\leq B,u \leq a} \big|\log\{1-\hat{F}_{n,\omega}^{\bm{\beta}}(u)\} - \Lambda_n(\bm{\beta},u)\big| = O_p(n^{-1/2+3\gamma+\varepsilon/2}),
    \label{hazard1}
\end{equation}

\begin{equation}
  \sup_{\|\bm{\beta}-\bm{\beta}^\prime\|\leq n^{-r}, u\leq a} \big|\log\{1-\hat{F}_{n,\omega}^{\bm{\beta}}(u)\} - \Lambda_n(\bm{\beta},u) -\log\{1-\hat{F}_{n,\omega}^{\bm{\beta}^\prime}(u)\} + \Lambda_n(\bm{\beta}^\prime,u) \big|=O_p(n^{-1/2-r/4+3\gamma+\varepsilon/2}),
    \label{hazard2}
\end{equation}
 for every $\varepsilon>0$, $0<\gamma<1$, and $0\leq r<1$. Here, $
    \Lambda_n(\bm{\beta},u) = - \int_{-\infty}^u \frac{\dif \mathbb{E}L_{n,\omega}(\bm{\beta},s)}{\mathbb{E}N_{n,\omega}(\bm{\beta},s)}$,
where  $N_{n,\omega}(\bm{\beta},s) = \sum_{i = 1}^n \sum_{k=1}^K \omega_i I_{\{e_{ik}(\bm{\beta})\geq s\}}$ and $L_{n,\omega}(\bm{\beta},s) = \sum_{i = 1}^n \sum_{k=1}^K \omega_i I_{\{e_{ik}(\bm{\beta})\leq s, \Delta_{ik}=1\}}$ are the  weighted observed failure process at risk processes. 
\end{lemma}

\begin{lemma}\label{lem:2}
Under conditions (C1) -- (C9), the following asymptotic linearity holds: 
\begin{equation}
     \bm{U}_{n,\omega}(\bm{\beta}) = \bm{U}_{n,\omega}(\bm{\beta}_0) -n \bm{A}_\omega(\bm{\beta}-\bm{\beta}_0)+o(n^{1/2}+n||\bm{\beta}-\bm{\beta}_0||), 
     \label{linearU}
\end{equation} uniformly over $||\bm{\beta}-\bm{\beta}_0|| \leq n^{-1/3}$, almost surely. 

\end{lemma}

\begin{lemma}
Under conditions (C1) -- (C9), given a $\sqrt{n}$-consistent estimator $\bm{b}$ of $\bm{\beta}$ satisfying that $\bm{b}-\bm{\beta}=O_p(n^{-1/2})$,  we have $n^{-1/2} U_{nj,\omega}(\bm{\beta}_0, \bm{b},\bm{\hat{\alpha}}(\bm{b})) = O_p(1), j=1,\dots,p,$ where $U_{nj,\omega}(\bm{\beta}_0, \bm{b},\bm{\hat{\alpha}}(\bm{b}))$ is $j$th element of $\bm{U}_{n,\omega}(\bm{\beta}_0, \bm{b}, \bm{\hat{\alpha}}(\bm{b}))$.
\label{lemma4}
\end{lemma}

\
\begin{proof}[Proof of Lemma 1]
It follows from the definition of $\hat{F}_{n,\omega}^{\bm{\beta}}(u)$ that, 
\begin{align}
    \log\{1-\hat{F}_{n,\omega}^{\bm{\beta}}(u)\} 
    &=- \int_{-\infty}^u \frac{\dif L_{n,\omega}(\bm{\beta},s)}{N_{n,\omega}(\bm{\beta},s)}-\int_{-\infty}^u O_p(N_{n,\omega}^{-2}(\bm{\beta},s))\dif L^\ast_{n,\omega}(\bm{\beta},s),
    \label{eq:TaylorKaplan}
\end{align}
where $L^\ast_{n,\omega}(\bm{\beta},s) =\sum_{i = 1}^n \sum_{k=1}^K \omega_i^2 I_{\{e_{ik}(\bm{\beta})\leq s, \Delta_{ik}=1\}}$. Notably, $\hat{F}_{n,\omega}^{\bm{\beta}}(u)$ involves potentially correlated error terms within each cluster and requires the additional effort to manage the weights.
Hence, the asymptotic properties derived based on the independence assumption of error terms in \citet{lai1988stochastic} cannot be applied directly. 
Define the $k$th weighted counting processes $N_{n,\omega}^{(k)}$ and $L_{n,\omega}^{(k)}$ by 
\begin{align}
    N_{n,\omega}^{(k)}(\bm{\beta},s) = \sum_{i = 1}^n \omega_i I_{\{e_{ik}(\bm{\beta})\geq s\}}
    ~~ \mbox{ and } ~~
    L_{n,\omega}^{(k)}(\bm{\beta},s) = \sum_{i = 1}^n \omega_i I_{\{e_{ik}(\bm{\beta})\leq s, \Delta_{ik}=1\}}, ~ k=1,\dots,K.
    \label{counting2}
\end{align}
For each fixed $k$, under condition (C1) and (C2), 
we make use of the independence of $e_{ik}, i=1,\dots,n$ and treat weights $\omega_i$ as bounded and covariate $x_i$ as nonrandom, it then follows from Theorem 2 in \citet{yang1997generalization} that, for any $\varepsilon>0$ and $0\leq r<1$, the weighted empirical processes defined in \eqref{counting2} satisfy
\begin{equation}
   \sup_{\|\bm{\beta}\|\leq B,s \leq a}\left|\mathbb{E}Q_{n,\omega}^{(k)}(\bm{\beta},s)-Q_{n,\omega}^{(k)}(\bm{\beta},s)\right|=o_p(n^{1/2+\varepsilon/2}),
    \label{w1}
\end{equation}
\begin{equation}
\sup_{\|\bm{\beta}-\bm{\beta}^\prime\| \leq n^{-r},s \leq a}\left|\mathbb{E}Q_{n,\omega}^{(k)}(\bm{\beta},s)-Q_{n,\omega}^{(k)}(\bm{\beta},s) - \mathbb{E}Q_{n,\omega}^{(k)}(\bm{\beta}^\prime,s)+ Q_{n,\omega}^{(k)}(\bm{\beta}^\prime,s) \right|=o_p(n^{1/2-r/4+\varepsilon/2}),
\label{w2}
\end{equation}
where $Q_{n,\omega}^{(k)}$ is either $N_{n,\omega}^{(k)}$ or $L_{n,\omega}^{(k)}$.
Moreover, by condition (C4) and continuity of error density $f$, we applied Lemma 2 in \citet{lai1988stochastic} and obtain that
that when $h \to 0$ and $nh \to \infty$, 
\begin{equation}
\sup_{\|\bm{\beta}-\bm{\beta}^\prime\|\leq h,s \leq a} |\mathbb{E}L_{n,\omega}^{(k)}(\bm{\beta},s) - \mathbb{E}L_{n,\omega}^{(k)}(\bm{\beta}^\prime,s)| = O(nh). 
    \label{w3}
\end{equation} 

Next, we define the ratio of counts and the ratio of expected counts for marginal error terms relative to pooled error terms. Utilizing the  properties for marginal counting processes established above, we bound the difference between $\log\{1-\hat{F}_{n,\omega}^{\bm{\beta}}(u)\}$ and $\Lambda_n(\bm{\beta},u)$ as follows. Let 
\begin{align*}
    & r_1^{(k)}(\bm{\beta},s) = \frac{\mathbb{E}N_{n,\omega}^{(k)}(\bm{\beta},s)}{\mathbb{E}N_{n,\omega}(\bm{\beta},s)}, 
    ~~ r_2^{(k)}(\bm{\beta},s) = \frac{N_{n,\omega}^{(k)}(\bm{\beta},s)}{N_{n,\omega}(\bm{\beta},s)}, 
    ~~ r_3^{(k)}(\bm{\beta}^\prime,s) = \frac{\mathbb{E}N_{n,\omega}^{(k)}(\bm{\beta}^\prime,s)}{\mathbb{E}N_{n,\omega}(\bm{\beta}^\prime,s)}, 
    ~~ \mbox{ and } 
    ~~ r_4^{(k)}(\bm{\beta}^\prime,s) = \frac{N_{n,\omega}^{(k)}(\bm{\beta}^\prime,s)}{N_{n,\omega}(\bm{\beta}^\prime,s)}.
\end{align*}
It follows from \eqref{eq:TaylorKaplan} and integration by parts that, 
\begin{align}
&\sup_{\|\bm{\beta}\|\leq B,u \leq a} \left|\log\{1-\hat{F}_{n,\omega}^{\bm{\beta}}(u)\}-\Lambda_n(\bm{\beta},u)\right| 
    \leq \sum_{k=1}^K \sup_{\|\bm{\beta}\|\leq B,u \leq a} 
    \left\{\int_{-\infty}^u  \left|\frac{r_1^{(k)}(\bm{\beta},s)}{\mathbb{E}N_{n,\omega}^{(k)}(\bm{\beta},s)} -\frac{r_2^{(k)}(\bm{\beta},s)}{N_{n,\omega}^{(k)}(\bm{\beta},s)}\right| \dif \mathbb{E}L_{n,\omega}^{(k)}(\bm{\beta},s)\right\} \notag \\
     &~~~~~~~~~~~~~ + 
    \sum_{k=1}^K \sup_{\|\bm{\beta}\|\leq B,u \leq a}\left\{ \left| \int_{-\infty}^u \frac{r_2^{(k)}(\bm{\beta},s)}{N_{n,\omega}^{(k)}(\bm{\beta},s)} \dif ( \mathbb{E}L_{n,\omega}^{(k)}(\bm{\beta},s) - L_{n,\omega}^{(k)}(\bm{\beta},s))  \right|  \right\} + \sup_{\|\bm{\beta}\|\leq B,u \leq a} \left|\sum_{k=1}^K \int_{-\infty}^u O_p(N_{n,\omega}^{-2}(\bm{\beta},s))\dif L^{(k)}_{n,\omega}(\bm{\beta},s)\right|.  
    \label{harzard3}
\end{align} 

To bound \eqref{harzard3}, we verify that the following order relations holds:
\begin{enumerate}[(i)]
\item $\sup_{\|\bm{\beta}\|\leq B,u \leq a} \{N_{n,\omega}^{(k)}(\bm{\beta},u)^{-1}\}=O_p(n^{-1})$.
Because $\sum_{i=1}^n n^{-2}\mathrm{Var}(I_{\{e_{ik}(\bm{\beta})\geq a\}})
< \infty$, it then follows from Kolmogorov's strong law of large number that $N_{n,\omega}^{(k)}(\bm{\beta}, a)/n - \sum_{i=1}^n  \omega_i \Pr(e_{ik}(\bm{\beta})\geq a)/n$ converges to zero almost surely as $n\to\infty$. 
Hence $N_{n,\omega}^{(k)}(\bm{\beta},a)/n = \sum_{i=1}^n  \omega_i \Pr(e_{ik}(\bm{\beta})\geq a)/n  +o_p(1)$ and $\{N_{n,\omega}^{(k)}(\bm{\beta},a)/n\}^{-1} = \{\sum_{i=1}^n  \omega_i \Pr(e_{ik}(\bm{\beta})\geq a)/n\}^{-1} + o_p(1)$ again because of the continuous mapping theorem and $\Pr(e_{ik}(\bm{\beta})\geq a) \geq \kappa_0 > 0$. Note that $\{N_{n,\omega}^{(k)}(\bm{\beta},u)\}^{-1} \leq \{N_{n,\omega}^{(k)}(\bm{\beta},a)\}^{-1}$ when $u\leq a$.
Let $M_0=(\kappa_0\Bar{W})^{-1}$, where $\bar{W}=\sum_{i=1}^n  \omega_i/n$ and $\omega_i$'s are uniformly bounded under conditions (6), we have 
\begin{equation}
    \begin{split}
\sup_{u\leq a} n\{N_{n,\omega}^{(k)}(\bm{\beta},u)\}^{-1} = n\{N_{n,\omega}^{(k)}(\bm{\beta},a)\}^{-1} = n\Big\{\sum_{i=1}^n  \omega_i \Pr(e_{ik}(\bm{\beta})\geq a)\Big\}^{-1}  + o_p(1) \leq M_0 + o_p(1). 
  \label{boundN}
    \end{split}
\end{equation}
Therefore, $\sup_{\|\bm{\beta}\|\leq B,u \leq a} \{|N_{n,\omega}^{(k)}(\bm{\beta},u)|^{-1}\}=O_p(n^{-1})$  because of condition (C5).  
\item $\sup_{\|\bm{\beta}\|\leq B}\left| \frac{1}{\mathbb{E}N_{n,\omega}^{(k)}(\bm{\beta},s)}-\frac{1}{N_{n,\omega}^{(k)}(\bm{\beta},s)}\right|
=o_p(n^{-3/2+3\gamma+\epsilon/2})$, for every $0<\gamma<1$, $\epsilon>0$.
We adopt the weighting function $p_n(x)=p(n^\gamma(x-cn^{-\gamma}))$ for tail modification defined in Theorem 3 in \citet{lai1988stochastic} but introduce a new $g_n(x)=\{(\sum_{i=1}^n \omega_i)^{3\gamma} p_n(x) x\}^{-1}$ on $(0,1)$. Given that weights $\omega_i$'s are bounded, it can be shown using the same arguments in the proof of Theorem 3 in \citet{lai1988stochastic} that 
\begin{equation}
\sup_{\|\bm{\beta}\|\leq B}\left|g_n(\frac{\mathbb{E}N_{n,\omega}^{(k)}(\bm{\beta},s)}{\sum_{i=1}^n \omega_i})-g_n(\frac{N_{n,\omega}^{(k)}(\bm{\beta},s)}{\sum_{i=1}^n \omega_i})\right|
=o_p(n^{-1/2+\varepsilon/2}). 
\label{g2}
\end{equation}
Hence,
\begin{equation}
    \sup_{\|\bm{\beta}\|\leq B}\left| \frac{1}{\mathbb{E}N_{n,\omega}^{(k)}(\bm{\beta},s)}-\frac{1}{N_{n,\omega}^{(k)}(\bm{\beta},s)}\right|\leq \sup_{\|\bm{\beta}\|\leq B}(\sum_{i=1}^n \omega_i)^{3\gamma-1}\left|g_n(\frac{\mathbb{E}N_{n,\omega}^{(k)}(\bm{\beta},s)}{\sum_{i=1}^n \omega_i})-g_n(\frac{N_{n,\omega}^{(k)}(\bm{\beta},s)}{\sum_{i=1}^n \omega_i})\right|=o_p(n^{-3/2+3\gamma+\epsilon/2}).
    \label{ii-g2}
\end{equation}

\item $\sup_{\|\bm{\beta}\|\leq B,s\leq a}\left|r_1^{(k)}(\bm{\beta},s)-r_2^{(k)}(\bm{\beta},s)\right|=O_p(n^{-1/2+3\gamma+\epsilon/2})$.
Making use of \eqref{ii-g2}, and $|r_1^{(k)}(\bm{\beta},s)| \leq1$, $|r_2^{(k)}(\bm{\beta},s)|\leq1$ for $k=1,\dots,K$, we have
\begin{equation}
     \sup_{\|\bm{\beta}\|\leq B,s\leq a} \left| \frac{1}{\mathbb{E}N_{n,\omega}(\bm{\beta},s)}-\frac{1}{N_{n,\omega}(\bm{\beta},s)} \right| \leq \sup_{\|\bm{\beta}\|\leq B,s\leq a} \left[\sum_{j=1}^K \left|  \left\{\frac{1}{\mathbb{E}N^{(j)}_{n,\omega}(\bm{\beta},s)}-\frac{1}{N^{(j)}_{n,\omega}(\bm{\beta},s)}\right\} r_1^{(j)}(\bm{\beta},s) r_2^{(j)}(\bm{\beta},s)\right|\right] = o_p(n^{-3/2+3\gamma+\epsilon/2}).
\end{equation}
Therefore,
\begin{align}
    &\sup_{\|\bm{\beta}\|\leq B,s\leq a}\left|r_1^{(k)}(\bm{\beta},s)-r_2^{(k)}(\bm{\beta},s)\right|= \sup_{\|\bm{\beta}\|\leq B,s\leq a}\left|  \frac{\mathbb{E}N_{n,\omega}^{(k)}(\bm{\beta},s)}{\mathbb{E}N_{n,\omega}(\bm{\beta},s)}-\frac{N_{n,\omega}^{(k)}(\bm{\beta},s)}{N_{n,\omega}(\bm{\beta},s)} \right| \notag\\
    \leq& \sup_{\|\bm{\beta}\|\leq B,s\leq a} \left[\left| \mathbb{E}N_{n,\omega}^{(k)}(\bm{\beta},s) \left\{\frac{1}{\mathbb{E}N_{n,\omega}(\bm{\beta},s)}-\frac{1}{N_{n,\omega}(\bm{\beta},s)}\right\} \right| + \left|\{\mathbb{E}N_{n,\omega}^{(k)}(\bm{\beta},s) - N_{n,\omega}^{(k)}(\bm{\beta},s)\} \frac{1}{N_{n,\omega}(\bm{\beta},s)} \right|\right] \notag \\
    \leq &O_p(n)o_p(n^{-3/2+3\gamma+\epsilon/2}) + o_p(n^{1/2+\epsilon/2})O_p(n^{-1}) = O_p(n^{-1/2+3\gamma+\epsilon/2}).
\end{align}
    
\item $\int_{-\infty}^u \dif \mathbb{E}L_{n,\omega}^{(k)}(\bm{\beta},s)= O_p(n)$. This is because $\int_{-\infty}^u \dif \mathbb{E}L_{n,\omega}^{(k)}(\bm{\beta},s)=\int_{-\infty}^u \sum_{i=1}^n \omega_i \Pr(C_{ik}-\bm{\beta}\bm{X}_{ik} \geq u)\dif F(u) \leq \sum_{i=1}^n \omega_i=O_p(n)$.

\end{enumerate}


Making use of (i)-(iii), we obtain
\begin{align}
        &\sup_{\|\bm{\beta}\|\leq B,s \leq a}\left|\frac{r_1^{(k)}(\bm{\beta},s)}{\mathbb{E}N_{n,\omega}^{(k)}(\bm{\beta},s)}\right. -\left.\frac{r_2^{(k)}(\bm{\beta},s)}{N_{n,\omega}^{(k)}(\bm{\beta},s)}\right| 
        \leq \sup_{\|\bm{\beta}\|\leq B,s \leq a}\left[ \left|   r_1^{(k)}(\bm{\beta},s) \left\{\frac{1}{\mathbb{E}N_{n,\omega}^{(k)}(\bm{\beta},s)}-\frac{1}{N_{n,\omega}^{(k)}(\bm{\beta},s)}\right\}\right| + \left|\{r_1^{(k)}(\bm{\beta},s)-r_2^{(k)}(\bm{\beta},s)\}\frac{1}{N_{n,\omega}^{(k)}(\bm{\beta},s)}\right|\right]\notag\\       
        &= O_p\left(n^{-3/2+3\gamma+\varepsilon/2}\right) + O_p(n^{-1/2+3\gamma+\epsilon/2})O_p(n^{-1})  
        =O_p(n^{-3/2+3\gamma+\varepsilon/2}), 
        \label{counting3}
\end{align}
   By further employing (iv), the first term of the right-hand side in \eqref{harzard3} can be bounded as follows. 
\begin{equation}
\sum_{k=1}^K\sup_{\|\bm{\beta}\|\leq B,u \leq a}
    \left\{\int_{-\infty}^u  \left|\frac{r_1^{(k)}(\bm{\beta},s)}{\mathbb{E}N_{n,\omega}^{(k)}(\bm{\beta},s)} -\frac{r_2^{(k)}(\bm{\beta},s)}{N_{n,\omega}^{(k)}(\bm{\beta},s)}\right| \dif \mathbb{E}L_{n,\omega}^{(k)}(\bm{\beta},s) \right\} = O_p\left(n^{-\frac{3}{2}+3\gamma+\frac{\epsilon}{2}}\right)O(n)=O_p\left(n^{-\frac{1}{2}+3\gamma+\frac{\epsilon}{2}}\right).
 \label{int1}
\end{equation}

The second term of the right-hand side in \eqref{harzard3} is bounded by,
\begin{align}
&\sum_{k=1}^K \sup_{\|\bm{\beta}\|\leq B,u \leq a}\left[ \left| \int_{-\infty}^u \frac{r_2^{(k)}(\bm{\beta},s)}{N_{n,\omega}^{(k)}(\bm{\beta},s)} \dif \{ \mathbb{E}L_{n,\omega}^{(k)}(\bm{\beta},s) - L_{n,\omega}^{(k)}(\bm{\beta},s)\}  \right|  \right]\notag\\
\leq&\sum_{k=1}^K\sup_{\|\bm{\beta}\|\leq B,u \leq a} \left\{\int_{-\infty}^u \left| \mathbb{E}L_{n,\omega}^{(k)}(\bm{\beta},s) - L_{n,\omega}^{(k)}(\bm{\beta},s) \right| \dif \frac{1}{N_{n,\omega}^{(k)}(\bm{\beta},s)}\right\}  + \sum_{k=1}^K\sup_{\|\bm{\beta}\|\leq B,u \leq a}\left\{\left| \mathbb{E}L_{n,\omega}^{(k)}(\bm{\beta},u) - L_{n,\omega}^{(k)}(\bm{\beta},u) \right| \frac{1}{N_{n,\omega}^{(k)}(\bm{\beta},s)}\right\}\notag\\
=& o_p\left(n^{\frac{1}{2}+\frac{\varepsilon}{2}}\right)O_p(1)=o_p\left(n^{\frac{1}{2}+\frac{\varepsilon}{2}}\right),
 \label{int2}
\end{align}
where 
the last equality follows from \eqref{w1} and $\sup_{\|\bm{\beta}\|\leq B,u \leq a}\int_{-\infty}^u   \dif \frac{1}{N_{n,\omega}^{(k)}(\bm{\beta},s)}=O_p(1)$  by definition of $N_{n,\omega}^{(k)}$.

It follows from boundedness (i) that $\sup_{\|\bm{\beta}\|\leq B,u \leq a} N_{n,\omega}^{-2}(\bm{\beta},s)=O_p(n^{-2})$. Further, by employing \eqref{w1} and (iv), the last term of \eqref{harzard3} can be bounded as follows,
\begin{align}
&\sup_{\|\bm{\beta}\|\leq B,u \leq a}  \left| \int_{-\infty}^u O_p(N_{n,\omega}^{-2}(\bm{\beta},s))\dif L^{(k)}_{n,\omega}(\bm{\beta},s)\right|
\\
\leq&\sum_{k=1}^K \sup_{\|\bm{\beta}\|\leq B,u \leq a} \left|\int_{-\infty}^u O_p(N_{n,\omega}^{-2}(\bm{\beta},s))\dif ( \mathbb{E}L_{n,\omega}^{(k)}(\bm{\beta},s) - L_{n,\omega}^{(k)}(\bm{\beta},s))\right| +  \sum_{k=1}^K \sup_{\|\bm{\beta}\|\leq B,u \leq a} \left|\int_{-\infty}^u O_p(N_{n,\omega}^{-2}(\bm{\beta},s))\dif  \mathbb{E}L_{n,\omega}^{(k)}(\bm{\beta},s)\right|\notag\\
=& O_p(n^{-2})o_p\left(n^{\frac{1}{2}+\frac{\varepsilon}{2}}\right) + O_p(n^{-2})O(\sum_{i=1}^n\omega_i)=O_p(n^{-1+\frac{\epsilon}{2}}).
\label{remining}
\end{align}
 Subsequently, \eqref{harzard3} can be bounded by 
\begin{align*}
    \sup_{\|\bm{\beta}\|\leq B,u \leq a} \left|\log\{1-\hat{F}_{n,\omega}^{\bm{\beta}}(u)\}-\Lambda_n(\bm{\beta},u)\right|
    =O_p\left(n^{-\frac{1}{2}+3\gamma+\frac{\epsilon}{2}}\right)+o_p\left(n^{\frac{1}{2}+\frac{\varepsilon}{2}}\right) +O_p(n^{-1+\frac{\epsilon}{2}})=O_p\left(n^{-\frac{1}{2}+3\gamma+\frac{\epsilon}{2}}\right),
\end{align*}
and \eqref{hazard1} is obtained. 

For \eqref{hazard2}, by setting $h=n^{-r}$ in  \eqref{w3}, we have
\begin{align}
    &\sup_{\|\bm{\beta}-\bm{\beta}^\prime\|\leq n^{-r},u \leq a} \left|\sum_{k=1}^K\int_{-\infty}^u  \left\{ \frac{r_4^{(k)}(\bm{\beta}^\prime,s)}{\mathbb{E}N_{n,\omega}^{(k)}(\bm{\beta}^\prime,s)}  \right\} \dif \{ \mathbb{E}L_{n,\omega}^{(k)}(\bm{\beta},s) - \mathbb{E}L_{n,\omega}^{(k)}(\bm{\beta}^\prime,s)\}\right|=O(n^{-r}),
\end{align}
and 
\begin{equation}
    \int_{-\infty}^u  \left( \frac{r_4^{(k)}(\bm{\beta}^\prime,s)}{N_{n,\omega}^{(k)}(\bm{\beta}^\prime,s)}  \right) \dif \{ \mathbb{E}L_{n,\omega}^{(k)}(\bm{\beta},s) - \mathbb{E}L_{n,\omega}^{(k)}(\bm{\beta}^\prime,s)\} = O_p(n^{-r})
\end{equation}
With a similar derivation of \eqref{harzard3} but applying property \eqref{w2} when $\|\bm{\beta}-\bm{\beta}^\prime\|\leq n^{-r}$, it can be shown that 
\begin{equation*}
 \sup_{\|\bm{\beta}-\bm{\beta}^\prime\|\leq n^{-r},u \leq a}\left|\int_{-\infty}^u \frac{\dif \mathbb{E}L_{n,\omega}(\bm{\beta},s)}{\mathbb{E}N_{n,\omega}(\bm{\beta},s)} - \int_{-\infty}^u \frac{\dif L_{n,\omega}(\bm{\beta},s)}{N_{n,\omega}(\bm{\beta},s)} - \int_{-\infty}^u \frac{\dif \mathbb{E}L_{n,\omega}(\bm{\beta}^\prime,s)}{\mathbb{E}N_{n,\omega}(\bm{\beta}^\prime,s)} + \int_{-\infty}^u \frac{\dif L_{n,\omega}(\bm{\beta}^\prime,s)}{N_{n,\omega}(\bm{\beta}^\prime,s)}\right|
     = O_p(n^{-1/2-r/4+3\gamma+\varepsilon/2}).
\end{equation*} 
Lemma 1 follows by further making use of \eqref{remining} for remainder terms.
\end{proof}

\begin{proof}[Proof of Lemma 2]
Similar with \citet{jin2006least}, to address the potential instabilities associated with the tail of $n^{-1}N_{n,\omega}(\bm{\beta},t)$, we adopt the tail modification $p_n(x)=p(n^\gamma(x-cn^{-\gamma}))$ introduced in \citet{lai1991large} when formulating the estimating functions.  With the same idea as in the proof of our Lemma 1, we express $\bm{U}_{n,\omega}(\bm{\beta}) $ in terms of marginal counting processes. 
Define 
\begin{align}
\bm{U}^{(k)}_{n,\omega}(\bm{\beta}) = - \int_{-\infty}^{\infty} t \,\dif p_n(n^{-1}N_{n,\omega}(\bm{\beta},t))\bm{Z}_{n,\omega}^{(k)}(\bm{\beta},t) -\int_{-\infty}^{\infty}  \frac{\int_{t}^\infty \{1-\hat{F}_{n,\omega}^{\bm{\beta}}(u)\}p_n(n^{-1}N_{n,\omega}(\bm{\beta},u)) \,\dif u}{1 - \hat{F}_{n,\omega}^{\bm{\beta}}(t)} 
 \,\dif \bm{J}^{(k)}_{n,\omega}(\bm{\beta},t),
\end{align}
\begin{align}
    \bm{\zeta}^{(k)}_{n,\omega}(\bm{\beta}) =  -\int_{-\infty}^{\infty} t \,\dif p_n(n^{-1}\mathbb{E}N_{n,\omega}(\bm{\beta},t))\mathbb{E}\bm{Z}^{(k)}_{n,\omega}(\bm{\beta},t) - \int_{-\infty}^{\infty}  \frac{\int_{t}^\infty \{1-F_{n,\omega}(\bm{\beta},u)\}p_n(n^{-1}\mathbb{E}N_{n,\omega}(\bm{\beta},u)) \,\dif u}{1 - F_{n,\omega}(\bm{\beta},t)}\,\dif \mathbb{E}\bm{J}^{(k)}_{n,\omega}(\bm{\beta},t),
\end{align}
where $Z^{(k)}_{nl,\omega}(\bm{\beta},t)=\sum_{i=1}^n \omega_i[(\bmX_i - \bar\bmX_\omega)^\top\bm{\Omega}^{-1}\{\bm{\alpha}(\bm{\beta})\}]_{lk} I_{\{e_{ik}(\bm{\beta})\geq t\}}$ and $ J_{nl,\omega}^{(k)}(\bm{\beta},t)=\sum_{k=1}^n \omega_i[(\bmX_i - \bar\bmX_\omega)^\top\bm{\Omega}^{-1}\{\bm{\alpha}(\bm{\beta})\}]_{lk}I_{\{e_{ik}(\bm{\beta})\geq t, \Delta_{ik}=0\}}$ are marginal weighted counting processes. Note that
 $\bm{U}_{n,\omega}(\bm{\beta}) 
      = \sum_{k=1}^K \bm{U}^{(k)}_{n,\omega}(\bm{\beta})$, and we define $\bm{\zeta}_{n,\omega}(\bm{\beta}) = \sum_{k=1}^K \bm{\zeta}_{n,\omega}^{(k)}(\bm{\beta})$.
Here, $F_{n,\omega}(\bm{\beta},u) = 1-\exp\left[-\int_{-\infty}^u \{1/\mathbb{E}N_{n,\omega}(\bm{\beta},u)\}\dif \mathbb{E}L_{n,\omega}(\bm{\beta},u)\right]$, 
and $F_{n,\omega}$ agree with $F$ when $\bm{\beta}=\bm{\beta}_0$.
We further prepare the closeness between $\hat{F}_{n,\omega}^{\bm{\beta}}(u)$  and  $F_{n,\omega}(\bm{\beta},u)$.
Following from Lemma \ref{lem:wKM} and the mean value theorem, we have that
for every $\varepsilon>0$, $0<r<1$ and $0<\gamma<1$,
\begin{equation}
    \sup_{\|\bm{\beta}\|\leq B,u \leq a} \big| \hat{F}_{n,\omega}^{\bm{\beta}}(u) -  F_{n,\omega}(\bm{\beta},u) \big| = O_p(n^{-1/2+3\gamma+\varepsilon/2}),
    \label{F1}
\end{equation}
\begin{equation}
    \sup_{\|\bm{\beta}-\bm{\beta}_0\|\leq n^{-r}, u\leq a} \big|\hat{F}_{n,\omega}^{\bm{\beta}}(u) -  F_{n,\omega}(\bm{\beta},u)  -\hat{F}_{n,\omega}^{\bm{\beta}_0}(u) + F(u) \big|=O_p(n^{-1/2-r/4+3\gamma+\varepsilon/2}).
    \label{F2}
\end{equation}
   For a fixed $k$, leveraging \eqref{F1}, \eqref{F2}, and the independence of $e_{ik}(\bm{\beta})$ among $i=1,\dots,n$, we apply Theorem 3 in \citet{lai1988stochastic} to obtain,
 for any $\varepsilon>0, 0<r<1$, 
\begin{equation}
    \sup_{\|\bm{\beta}-\bm{\beta}_0\|\leq n^{-r}
    } \left|\bm{U}^{(k)}_{n,\omega}(\bm{\beta}) -\bm{U}^{(k)}_{n,\omega}(\bm{\beta}_0) -\bm{\zeta}^{(k)}_{n,\omega}(\bm{\beta}) +\bm{\zeta}^{(k)}_{n,\omega}(\bm{\beta}_0) \right|=o_p(n^{1/2-r/4+3\gamma+\varepsilon/2}). 
    \label{nonran0}
\end{equation}
Therefore, 
we combine \eqref{nonran0} for all $k$ and obtain that
\begin{equation}
    \sup_{\|\bm{\beta}-\bm{\beta}_0\|\leq n^{-r}} \left|\bm{U}_{n,\omega}(\bm{\beta}) -\bm{U}_{n,\omega}(\bm{\beta}_0) -\bm{\zeta}_{n,\omega}(\bm{\beta}) +\bm{\zeta}_{n,\omega}(\bm{\beta}_0) \right|=o_p(n^{1/2-r/4+3\gamma+\varepsilon/2}).
    \label{nonran}
\end{equation}

We next show that $\bm{\zeta}_{n,\omega}(\bm{\beta})= -n(\bm{\beta}-\bm{\beta}_0)^\top \bm{A}_\omega + o(n\|\bm{\beta}-\bm{\beta}_0\|)$ when $\|\bm{\beta}-\bm{\beta}_0\|\leq n^{-r/2}$. 
Using the similar arguments in the proof of Lemma 1 in \citet{lai1991large}, it can be shown that, as $\bm{\beta}$ is approaching $\bm{\beta}_0$, the $l$th element of $\bm{\zeta}_{n,\omega}(\bm{\beta})$ has the following form after simplification,  
\begin{align}
   \zeta_{nl,\omega}(\bm{\beta})
    &= - n(\bm{\beta}-\bm{\beta}_0)^\top \int_{-\infty}^{\infty} \psi(t)  \left\{ n^{-1}\sum_{i = 1}^n \sum_{j=1}^K   \left[   \omega_i (\bmX_{ij})^\top [(\bmX_i - \bar\bmX_\omega)^\top\bm{\Omega}^{-1}\{\bm{\alpha}(\bm{\beta})\}]_{lj}G_{ij}^{\bm{\beta}}(t)f(t) \right]\dif t \right. \notag\\
    &~~~~~~- \left.  \left[ n^{-1}\sum_{i = 1}^n \sum_{j=1}^K  \omega_i  [(\bmX_i - \bar\bmX_\omega)^\top\bm{\Omega}^{-1}\{\bm{\alpha}(\bm{\beta})\}]_{lj}G_{ij}^{\bm{\beta}}(t) \right] \frac{\sum_{i = 1}^n \sum_{j=1}^K \{\bmX_{ij}^\top G_{ij}^{\bm{\beta}}(t)\}}{\sum_{i = 1}^n \sum_{j=1}^K G_{ij}^{\bm{\beta}}(t)} f(t) \dif t  \right\} + 
    o(n\|\bm{\beta}-\bm{\beta}_0\|), 
\end{align}
where $\psi(t)=\int_{t}^s\{1-F(s)\}\dif s [f^\prime(t)/f(t)+f(t)/\{1-F(t)\}]/\{1-F(t)\}$, and $G_{ij}^{\bm{\beta}}(t)=\Pr(C_{ij}-\bmX_{ij}\bm{\beta}\geq t)$. 

With condition (C9), we obtain from the dominated convergent theorem that 
$\zeta_{nl,\omega}(\bm{\beta})=-n\bm{A}_{\omega,\ell}(\bm{\beta}-\bm{\beta}_0) + o(n\|\bm{\beta}-\bm{\beta}_0)\|)$, where $\bm{A}_{\omega,\ell}$ is $l$th row of $\bm{A}_\omega$ which is defined as 
\begin{equation}
    (\bm{A}_{\omega,\ell})^{\top} = \int_{-\infty}^{\infty} \left[ \left\{\bm{\Gamma}^{\omega,\ell}_2(t)-\frac{\bm{\Gamma}_1(t)\Gamma^{\omega,\ell}_1(t)}{\Gamma_0(t)}\right\} \int_t^\infty \{1-F(s)\}\, ds \right]\,d\lambda(t).
\end{equation}
Combining this with statement \eqref{nonran}, Lemma 2 follows.
\end{proof}

\begin{proof}[Proof of Lemma 3]
Define $\bm{B}_{n,\omega}(\bm{b})=\sum_{i = 1}^n (\bmX_i - \bar\bmX_\omega)^\top\omega_i \bm{\Omega}^{-1}(\bm{\hat{\alpha}}(\bm{b}))\bmX_i(\bm{b} - \bm{\beta}_0)$ and $\bm{\bar{B}}_{n,\omega}(\bm{b})=\sum_{i = 1}^n(\bmX_i - \bar\bmX_\omega)^\top\omega_i\bar{\bm{\Omega}}^{-1}\bmX_i(\bm{b} - \bm{\beta}_0)$,  it follows from Lemma \ref{lem:2} that, uniformly in $||\bm{b}-\bm{\beta}_0\|\leq n^{-1/3}$, 
\begin{align}
 &\bm{U}_{n,\omega}(\bm{\beta}_0, \bm{b}, \bm{\hat{\alpha}}(\bm{b})) 
     = \bm{U}_{n,\omega}(\bm{b})+ \sum_{i = 1}^n(\bmX_i - \bar\bmX_\omega)^\top\omega_i\bm{\Omega}^{-1}(\bm{\hat{\alpha}}(\bm{b}))\bmX_i(\bm{b} - \bm{\beta}_0) \notag\\
    & =\bm{U}_{n,\omega}(\bm{\beta}_0) - n\bm{A}_\omega(\bm{b}-\bm{\beta}_0) + \bar{\bm{B}}_{n,\omega}(\bm{b}) + \{\bm{B}_{n,\omega}(\bm{b}) - \bar{\bm{B}}_{n,\omega}(\bm{b})\} + o(n^{1/2} + n||\bm{b}-\bm{\beta}_0\|).
\end{align}
Hence, $U_{nj,\omega}(\bm{\beta}_0, \bm{b}, \bm{\hat{\alpha}}(\bm{b}))=U_{nj,\omega}(\bm{\beta}_0) - n\bm{A}_{\omega,j}(\bm{b}-\bm{\beta}_0) + \bar{B}_{nj,\omega}(\bm{b}) + \{B_{nj,\omega}(\bm{b}) -\bar{B}_{nj,\omega}(\bm{b})\}$, where $\bm{A}_{\omega,j}$ is $j$th row of matrix $\bm{A}_\omega$, $\bar{B}_{nj,\omega}(\bm{b})$  and $B_{nj,\omega}(\bm{b})$ are the $j$th element of $\bar{\bm{B}}_{n,\omega}(\bm{b})$ and $\bm{B}_{n,\omega}(\bm{b})$, respectively.  
Since $||\bm{b}-\bm{\beta}_0\| = O_p(n^{-1/2})$ and $\bm{A}_{\omega,j}$ is nonrandom, we have $||n\bm{A}_{\omega,j}(\bm{b}-\bm{\beta}_0)\|=O_p(n^{1/2})$. 

We next show that $B_{nj,\omega}(\bm{b}) -\bar{B}_{nj,\omega}(\bm{b}) = O_p(1)$ and $\bar{B}_j= O_p(n^{1/2})$.
Under condition (C8),  
\begin{align}
     ||B_{nj,\omega}(\bm{b}) -\bar{B}_{nj,\omega}(\bm{b})|| 
     &= ||\sum_{i = 1}^n (\bmX_{ij} - \bar\bmX_{\omega,j})^\top\omega_i \{\bm{\Omega}^{-1}(\bm{\hat{\alpha}}(\bm{b}))-\bar{\bm{\Omega}}^{-1}\}\bmX_i (\bm{b} - \bm{\beta}_0)|| \notag\\
    &\leq \sum_{i = 1}^n  ||(\bmX_{ij} - \bar\bmX_{\omega,j})^\top||_F \cdot ||\bm{\Omega}^{-1}(\bm{\hat{\alpha}}(\bm{b}))-\bar{\bm{\Omega}}^{-1}|| \cdot ||\bmX_i|| \cdot ||\bm{b} - \bm{\beta}_0|| \notag\\
    & = nO_p(n^{-1/2})O_p(n^{-1/2}) = O_p(1),
\end{align}
where $\bar\bmX_{\omega,j}$ is the $j$th row of $\bar\bmX_{\omega}$.
Moreover, since $\bmX_{i}$ are bounded uniformly and nonrandom,  $\bar{\bm{\Omega}}^{-1}$ is a finite constant positive definite matrix and $\omega_i$ are finite and does not involve in any parameter, we have, with probability tending to 1,
$
 \bar{B}_j=\sum_{i = 1}^n [(\bmX_i - \bar\bmX_\omega)^\top\omega_i\bar{\bm{\Omega}}^{-1}\bmX_i]_j (\bm{b} - \bm{\beta}_0)=nO_p(n^{-1/2})=O_p(n^{1/2})$, where
$[(\bmX_i - \bar\bmX_\omega)^\top\omega_i\bar{\bm{\Omega}}^{-1}\bmX_i]_j$ is the $j$th row of matrix $(\bmX_i - \bar\bmX_\omega)^\top\omega_i\bar{\bm{\Omega}}^{-1}\bmX_i$. 
Therefore, Lemma 3 follows.
\end{proof}

\begin{proof}[\textbf{Proof of  Theorem 1}]

Define the penalized objective function of $\tilde{\bm{U}}_n^\omega(\bm{\beta}, \bm{b}, \bm{\hat{\alpha}}(\bm{b})))$ as
\begin{equation}
    \tilde{S}_{n,\omega}(\bm{\beta}, \bm{b}, \bm{\hat{\alpha}}(\bm{b})) = -\sum_{i = 1}^n\left\{\hat\bmY_{i, \omega}(\bm{b}) - \bmX_i\bm{\beta}\right\}^\top\omega_i\bm{\Omega}^{-1}(\bm{\hat{\alpha}}(\bm{b}))\left\{\hat\bmY_{i, \omega}(\bm{b}) - \bmX_i\bm{\beta}\right\} - n\sum_{j=1}^p p_{\lambda_n}(|\beta_j|),
\end{equation}
such that $\partial \tilde{S}_{n,\omega}(\bm{\beta}, \bm{b}, \bm{\hat{\alpha}}(\bm{b})) /\partial \bm{\beta}= \tilde{\bm{U}}_{n,\omega}(\bm{\beta}, \bm{b}, \bm{\hat{\alpha}}(\bm{b}))$.
Given a fixed $\bm{b}$, finding the solution of $\tilde{\bm{U}}_n^\omega(\bm{\beta}, \bm{b}, \bm{\hat{\alpha}}(\bm{b}))=0$ with respect to $\bm{\beta}$ is equivalent to maximizing its objective function $\tilde{S}_{n,\omega}(\bm{\beta}, \bm{b}, \bm{\hat{\alpha}}(\bm{b}))$ with respect to $\bm{\beta}$. 
Recall that $a_n = n^{-1/2} + \max\{p^\prime_{\lambda_n}(|\beta_{j0}|):\beta_{j0} \neq 0 \}$.
Following the same idea in the proof of Theorem 1 in \citet{fan2001variable}, we wish to prove that, given a fixed $\bm{b}$, there exists a local minimizer $\hat{\bm{\beta}}$ of $\tilde{S}_{n,\omega}(\bm{\beta}, \bm{b}, \bm{\hat{\alpha}}(\bm{b}))$ such that $||\hat{\bm{\beta}} - \bm{\beta}_0 || = O_p(a_n)$.

With $C>0$, $p_{\lambda_n}(|\beta_{j0}|) = p_{\lambda_n}(0) = 0$ for $s<j\leq p$. For any $p \times 1$ vector $\bm{u}$ such that $||\bm{u}|| \leq C$,
\begin{align}
    & ~~~ \tilde{S}_{n,\omega}(\bm{\beta}_0 + a_n\bm{u}, \bm{b}, \bm{\hat{\alpha}}(\bm{b})) - \tilde{S}_{n,\omega} (\bm{\beta}_0, \bm{b}, \bm{\hat{\alpha}}(\bm{b})) \notag\\
    & \leq a_n\{\bm{U}_{n,\omega}(\bm{\beta}_0, \bm{b}, \bm{\hat{\alpha}}(\bm{b})\}^\top \bm{u} -  \frac{1}{2}a_n^2 \bm{u}^\top \bm{H}_{n,\omega} (\bm{b}) \bm{u}- \sum_{n=1}^s[na_np^\prime_{\lambda_n}(|\beta_j|)sgn(\beta_{j})u_j + na_n^2p^{\prime \prime}_{\lambda_n}(|\beta_j|)u_j^2\{1+o(1)\}],
\end{align}
where $S_{n,\omega}(\bm{\beta}, \bm{b},\bm{\hat{\alpha}}(\bm{b})) =-\sum_{i = 1}^n\left\{\hat\bmY_{i, \omega}(\bm{b}) - \bmX_i\bm{\beta}\right\}^\top\omega_i\bm{\Omega}^{-1}(\bm{\hat{\alpha}}(\bm{b}))\left\{\hat\bmY_{i, \omega}(\bm{b}) - \bmX_i\bm{\beta}\right\}$ is the unpenalized objective function, and $\bm{H}_{n,\omega} (\bm{b}) = \sum_{i=1}^n \bmX_i^\mathrm{T} \omega_i \bm{\Omega}^{-1}(\bm{\hat{\alpha}}(\bm{b}))(\bmX_i -  \bar{\bmX}_\omega)$. The last inequality is obtained from the Taylor expansion of the objective function. 
From Lemma \ref{lemma4}, $n^{-1/2}\bm{U}_{n,\omega}(\bm{\beta}_0, \bm{b},\bm{\hat{\alpha}}(\bm{b})) = O_p(1)$ given a $\sqrt{n}$-consistent estimator $\bm{b}$, and  $\bm{D}=\lim_{n \to \infty} \frac{1}{n} \bm{H}_{n,\omega} (\bm{b})$ exists and positive definite. 
Given that $\|\bm{u}\|$ is bounded by $C$, and $D$ is positive definite with finite eigenvalues, then $2a_n\bm{U}_{n,\omega}(\bm{\beta}_0, \bm{b}, \bm{\hat{\alpha}}(\bm{b}))^\top \bm{u} = O_p(n^{1/2}a_n)$ and $n a_n^2 \bm{u}^\top \bm{D} \bm{u} = O_p(na_n^2)$. Under the assumption that $\lambda_n \to 0$ as $n \to \infty$, we have $\max\{p^\prime_{\lambda_n}(|\beta_{j0}|):\beta_{j0} \neq 0 \} = 0$ as $n \to \infty$. Thus, $O_p(n^{1/2}a_n) = O_p(na_n^2)$. By choosing a large enough $C$, the second term on the right-hand side dominates the first term uniformly in $||\bm{u}||=C$. The third term is bounded by 
\begin{equation}
    \sqrt{s}na_n \max\{p^\prime_{\lambda_n}(|\beta_{j0}|):\beta_{j0} \neq 0 \} ||\bm{u}|| + n a_n^2\max\{p^{\prime \prime}_{\lambda_n}(|\beta_{j0}|):\beta_{j0} \neq 0 \} ||\bm{u}||^2,
\end{equation}
which is also dominated by the second term. 
 Then we have $\tilde{S}_{n,\omega}(\bm{\beta}_0 + a_n\bm{u}, \bm{b}, \bm{\hat{\alpha}}(\bm{b})) - \tilde{S}_{n,\omega} (\bm{\beta}_0, \bm{b}, \bm{\hat{\alpha}}(\bm{b})) < 0$ with a sufficiently large $C$. 
\end{proof}

\begin{proof}[\textbf{Proof of Theorem 2 (1)}]
Given $\bm{b}$, for any $\bm{\beta}$ satisfying $\bm{\beta}-\bm{\beta}_0=O_p(n^{-1/2})$ and $j=1,\dots,p$,
\begin{align}
    \frac{\partial \tilde{S}_{n,\omega} (\bm{\beta}, \bm{b}, \bm{\hat{\alpha}}(\bm{b}))}{\partial \beta_j} &= \frac{\partial S_{n,\omega} (\bm{\beta}, \bm{b}, \bm{\hat{\alpha}}(\bm{b}))}{\partial \beta_j} - n p^\prime_{\lambda_n}(|\beta_j|)sgn(\beta_{j})\notag\\
    & = U_{nj}^\omega(\bm{\beta}_0, \bm{b}, \bm{\hat{\alpha}}(\bm{b})) + \sum_{l=1}^p \frac{\partial \bm{U}_{n,\omega}(\bm{\beta}^\ast, \bm{b}, \bm{\hat{\alpha}}(\bm{b}))}{\partial \beta_l}(\beta_{l} - \beta_{l0}) - n p^\prime_{\lambda_n}(|\beta_j|)sgn(\beta_{j}) \notag\\
    & = O_p(n^{1/2}) -n p^\prime_{\lambda_n}(|\beta_j|)sgn(\beta_{j}) \notag\\
    &=-n \lambda_n (\frac{p^\prime_{\lambda_n}(|\beta_j|)}{\lambda_n}sgn(\beta_{j}) + O_p(n^{1/2}/\lambda_n)),
\end{align}
where $\bm{\beta}^\ast$ lies between $\bm{\beta}$ and $\bm{\beta}_0$. The third equality follows from Lemma \ref{lemma4} that $U^\omega_{nj}(\bm{\beta}_0,\bm{b},\bm{\alpha})=O_p(n^{1/2})$.

Given that $\bm{\beta}-\bm{\beta}_0=O_p(n^{-1/2})$, it is reasonable to assume that, for $j=d+1,\dots,p$, $|\beta_j|<Cn^{-1/2}$ for any constant $C$. Since $\liminf_{n \to \infty} \liminf_{\beta \to 0+}  p^\prime_{\lambda_n}(\beta)/\lambda_n>0$, we have
\begin{align}
    & \frac{\partial \tilde{S}_{n,\omega} (\bm{\beta}, \bm{b}, \bm{\hat{\alpha}}(\bm{b}))}{\partial \beta_j} <0, \mbox{ for } 0<\beta_j<Cn^{-1/2}, \mbox{ and } \notag\\
    & \frac{\partial \tilde{S}_{n,\omega} (\bm{\beta}, \bm{b}, \bm{\hat{\alpha}}(\bm{b}))}{\partial \beta_j} >0, \mbox{ for }  -Cn^{-1/2}<\beta_j<0, 
\end{align}
where $j=d+1,\dots,p$. Therefore, with probability tending to 1, for any approximated solution $\bm{\beta}$ such that 
 $\bm{\beta}_1$ satisfying $\bm{\beta}_1-\bm{\beta}_{01}=O_p(n^{-1/2})$, we further have $\bm{\beta}_2=\bm{0}$ as $n \to \infty$. 
~\\
\end{proof}

\begin{proof}[\textbf{Proof of Theorem 2 (2)}] Note that SCAD function satisfies that $\lim_{\beta \to 0+} p^\prime_{\lambda_n}(\beta) \leq \lambda_n$. Thus, for $j=d+1,\dots,p$,
\begin{align}
    |n^{-1}\tilde{U}_{nj,\omega}(\bm{\beta},\bm{b},\bm{\alpha})| =|-  p^\prime_{\lambda_n}(|\beta_j|)sgn(\beta_{j}) + O_p(n^{-1/2})|\leq \lambda_n + O_p(n^{-1/2}).
\end{align}
Under the assumption that $n^{1/2}\lambda_n \to \infty$, we have $|n^{-1}\tilde{U}_{nj,\omega}(\bm{\beta})| \leq \lambda_n, j=s+1,\dots,p.$

For $j=1,\dots,d$, it was shown by \cite{wang2012penalized} that $\Pr(|\beta_j| \geq a\lambda_n) \to 1$ implies that SCAD penalty term $p^\prime_{\lambda_n}(|\beta_j|) = 0$ with probability tending to 1. We have shown in the proof of (1) that an approximated solution $\hat{\bm{\beta}}$ must satisfy that $\Pr(\hat{\bm{\beta}}_2=\bm{0}) \to 1$, then $\hat{\bm{\beta}}$ can be written as $(\hat{\bm{\beta}}_1^\top, \bm{0}^\top)^\top$, which is the oracle estimator. We thus have $U^\omega_{nj}(\bm{\beta},\bm{b},\bm{\alpha})=0, j=1,\dots,d$. Therefore, as $n \to \infty$, 
\begin{align}
    |\tilde{U}_{nj,\omega}(\bm{\beta},\bm{b},\bm{\alpha})| &= |U^\omega_{nj}(\bm{\beta},\bm{b},\bm{\alpha}) -n p^\prime_{\lambda_n}(|\hat{\beta}_j|)sgn(\hat{\beta}_{j})|=0,
\end{align}
 with probability tending to 1.
\end{proof}

\begin{proof}[\textbf{Proof of Theorem 2 (3)}]
We follow the Theorem 2 of \citet{lai1991large} and obtain that, $n^{-1/2}\bm{U}_{n,\omega}(\bm{\beta_{0}})$ converges in distribution to a normal random variable with mean $\bm{0}$ and variance-covariance matrix $\bm{B}_{\omega}$ such that its $l$th row vector is given by 
\begin{equation}
    \bm{B}_{\omega,\ell}^\top=\int_{-\infty}^{\infty} \left [ \left(\bm{\Gamma}^{\omega,\ell}_2(t)-\frac{\bm{\Gamma}_1(t)\Gamma^{\omega,\ell}_1(t)}{\bm{\Gamma}_0(t)}\right) \frac{(\int_t^\infty \{1-F(s)\}\, ds)^2}{1-F(t)} \right]\ d\lambda(t).
\end{equation}
Given the asymptotical linearity of $\bm{U}_{n,\omega}(\bm{b})$ from Lemma \ref{lem:2}, we have, 
uniformly in $||\bm{b}-\bm{\beta}_0|| \leq n^{-1/3}$,
 \begin{align}     n^{-1/2}\bm{\tilde{U}}_{n,\omega}(\hat{\bm{\beta}},\bm{b},\bm{\alpha})
&=n^{-1/2}\bm{U}_{n,\omega}(\bm{b})+n^{-1/2}\sum_{i = 1}^n(\bmX_i - \bar\bmX_\omega)^\top\omega_i\bm{\Omega}^{-1}(\bm{\hat{\alpha}}(\bm{b}))\bmX_i(\bm{b} - \hat{\bm{\beta}})+\sqrt{n}\bm{q}_{\lambda_n}(\hat{\bm{\beta}}) \notag\\
&= n^{-1/2}\bm{U}_{n,\omega}(\bm{\beta}_0) -\sqrt{n} \bm{A}_\omega(\bm{b}-\bm{\beta}_0)+ \sqrt{n}\bm{D}_\omega(\bm{b})(\bm{b}-\hat{\bm{\beta}})+\sqrt{n}\bm{q}_{\lambda_n}(\hat{\bm{\beta}})+ o_p(1),
     \label{linearPU}
 \end{align}
 where $\bm{D}_\omega(\bm{b})=n^{-1}\sum_{i = 1}^n(\bmX_i - \bar\bmX_\omega)^\top\omega_i\bm{\Omega}^{-1}(\bm{\hat{\alpha}}(\bm{b}))\bmX_i$. 
  Under condition (C8),
\begin{align}
    \bm{D}_\omega(\bm{b}) = n^{-1}\sum_{i = 1}^n(\bmX_i - \bar\bmX_\omega)^\top\omega_i[\bm{\Omega}^{-1}\{\bm{\hat{\alpha}} (\bm{b})\}- \bar{\bm{\Omega}}]\bmX_i + n^{-1}\sum_{i = 1}^n(\bmX_i - \bar\bmX_\omega)^\top\omega_i \bar{\bm{\Omega}}\bmX_i.
\end{align}
Given that $\bm{\Omega}^{-1}\{\bm{\hat{\alpha}} (\bm{b})\}- \bar{\bm{\Omega}}=O_p(n^{-1/2})$, as $n \to \infty$, we have $ n^{-1}\sum_{i = 1}^n(\bmX_i - \bar\bmX_\omega)^\top\omega_i[\bm{\Omega}^{-1}\{\bm{\hat{\alpha}} (\bm{b})\}- \bar{\bm{\Omega}}]\bmX_i =O_p(n^{-1/2})$.  Therefore, 
  $\lim_{n \to \infty}\bm{D}_\omega(\bm{b})=\lim_{n \to \infty} n^{-1}\sum_{i = 1}^n(\bmX_i - \bar\bmX_\omega)^\top\omega_i \bar{\bm{\Omega}}\bmX_i$. This suggests that the limiting value exists, given that $\omega_i$ is  bounded and $\bar{\bm{\Omega}}$ is a constant matrix. We define $\bm{D}_\omega=\lim_{n \to \infty}\bm{D}_\omega(\bm{b})$. Let $\bm{\tilde{U}}^\ast_{n,\omega}(\cdot)=( \tilde{U}_{n1,\omega}(\cdot),\dots,\tilde{U}_{nd,\omega}(\cdot))^\top$ and $\bm{U}^\ast_{n,\omega}(\cdot)=( U_{n1,\omega}(\cdot),\dots,U_{nd,\omega}(\cdot))^\top$ denote the first $d$ elements of $\bm{\tilde{U}}_{n,\omega}(\cdot)$ and $\bm{U}_{n,\omega}(\cdot)$, respectively.
 Applying the Taylor expansion to the first $d$ elements of penalty term $\bm{q}_{\lambda_n}(\hat{\bm{\beta}})$, we obtain that, as $n \to \infty$,
 \begin{align}
     &n^{-1/2}\bm{\tilde{U}}_{n,\omega}^\ast (\hat{\bm{\beta}},\bm{b},\bm{\alpha}) \notag \\
     =&n^{-1/2}\bm{U}^\ast_{n,\omega}(\bm{\beta_{0}}) -\sqrt{n} \bm{A}_{11}^\omega(\bm{b}_{1}-\bm{\beta_{10}})+\sqrt{n}D_{11}^\omega(\bm{b}_1-\hat{\bm{\beta}}_1)+ \sqrt{n}\bm{q}_{01}+ \sqrt{n}\bm{\Lambda}_{11}(\hat{\bm{\beta}}_1-\bm{\beta_{10}}) +o_p(1) \notag\\     =&n^{-1/2}\bm{U}^\ast_{n,\omega}(\bm{\beta_{0}}) -\sqrt{n} (\bm{A}_{11}^\omega-\bm{\Lambda}_{11})\left[\hat{\bm{\beta}}_{1}-\bm{\beta_{10}} - (\bm{A}_{11}^\omega-\bm{\Lambda}_{11})^{-1}\bm{q}_{01}\right]-\sqrt{n}(\bm{A}_{11}^\omega-\bm{D}_{11}^\omega)(\bm{b}_1-\hat{\bm{\beta}}_{1})  +o_p(1),
 \end{align}
where $\bm{b}_1$ is the first $d$ elements of $\bm{b}$, $\bm{q}_{01}=\{p^\prime_{\lambda_n}(|{\beta_{10}|})\cdot \sgn(\beta_{10}),\dots,p^\prime_{\lambda_n}(|{\beta_{d0}|})\cdot \sgn(\beta_{d0})\}^\top$, $\bm{\Lambda}_{11}=diag(p^{\prime\prime}(|\beta_{10}|), \dots, p^{\prime\prime}(|\beta_{d0}|))$, 
and $\bm{A}_{11}^\omega$  and $\bm{D}_{11}^\omega$ are the first $d\times d$ sub-matrix of $\bm{A}_\omega$ and $\bm{D}_\omega$, respectively. 

In Theorem 2(2), we have shown that with probability tending to 1, $|\tilde{U}_{nj,\omega}(\hat{\bm{\beta}},\bm{b},\bm{\alpha})|=0$, for $j=1,\dots,d$,
as $n \to \infty$. Therefore, 
\begin{align}
    \sqrt{n}(\bm{A}_{11}^\omega-\bm{\Lambda}_{11})\left[\hat{\bm{\beta}}_1-\bm{\beta}_{10}-(\bm{A}_{11}^\omega-\bm{\Lambda}_{11})^{-1}\bm{q}_{01}  \right] + \sqrt{n}(\bm{A}_{11}^\omega-\bm{D}_{11}^\omega)(\bm{b}_1-\hat{\bm{\beta}}_{1})=n^{-1/2}\bm{U}_{n,\omega}(\bm{\beta_{10}})+o_p(1). 
\end{align}

Thus, we conclude that, given a root-n consistent and asymptotically normal distributed $\bm{b}$,
\begin{equation}
    \sqrt{n}(\bm{A}_{11}^\omega-\bm{\Lambda}_{11})\left[\hat{\bm{\beta}}_1-\bm{\beta}_{10}-(\bm{A}_{11}^\omega-\bm{\Lambda}_{11})^{-1}\bm{q}_{01} \right] + \sqrt{n}(\bm{A}_{11}^\omega-\bm{D}_{11}^\omega)(\bm{b}_1-\hat{\bm{\beta}}_{1}) \to _d N(\bm{0}, \bm{B}_{11}^\omega),
\end{equation}
where $\bm{A}_{11}^\omega$, $\bm{B}_{11}^\omega$, and  $\bm{D}_{11}^\omega$ are the first $d\times d$ sub-matrix of $\bm{A}_\omega$, $\bm{B}_\omega$, and  $\bm{D}_\omega$, respectively.

\end{proof}


\end{document}